\documentclass[final,5p,times,twocolumn]{elsarticle}

\usepackage{amssymb}
\usepackage{tabularx}
\usepackage{longtable}
\usepackage{xltabular}
\usepackage{soul}
\usepackage{amsmath}
\usepackage{rotating}
\usepackage{longtable}
\usepackage{pdflscape}
\usepackage{bbding}
\usepackage{multirow}
\usepackage{xcolor}
\usepackage{hyperref}
\usepackage{caption}
\usepackage{comment}
\usepackage{subcaption}
\usepackage{adjustbox}
\usepackage{array}
\usepackage{booktabs}
\usepackage{multirow}

\journal{Journal of Information Security and Application}

\begin{document}

\begin{frontmatter}



\title{Through the Static: Demystifying Malware Visualization via Explainability}


\author[inst1]{Matteo Brosolo\corref{cor1}}
\ead{matteo.brosolo@unipd.it}
\author[inst1]{Vinod P.\corref{cor1}}
\ead{vinod.puthuvath@unipd.it}
\author[inst1]{Mauro Conti}
\cortext[cor1]{Corresponding author}

\affiliation[inst1]{organization={Department of Mathematics, University of Padova},
            addressline={Via Trieste, 63}, 
            city={Padova},
            country={Italy}}

\begin{abstract}
Security researchers grapple with the surge of malicious files, necessitating swift identification and classification of malware strains for effective protection. Visual classifiers and in particular Convolutional Neural Networks~(CNNs) have emerged as vital tools for this task. However, issues of robustness and explainability, common in other high risk domain like medicine and autonomous vehicles, remain understudied in current literature. Although deep learning visualization classifiers presented in research obtain great results without the need for expert feature extraction, they have not been properly studied in terms of their replicability. Additionally, the literature is not clear on how these types of classifiers arrive to their answers. Our study addresses these gaps by replicating six CNN models and exploring their pitfalls. We employ Class Activation Maps~(CAMs), like GradCAM and HiResCAM, to assess model explainability. We evaluate the CNNs’ performance and interpretability on two standard datasets, MalImg and Big2015, and a newly created called VX-Zoo. We employ these different CAM techniques to gauge the explainability of each of the models. With these tools, we investigate the underlying factors contributing to different interpretations of inputs across the different models, empowering human researchers to discern patterns crucial for identifying distinct malware families and explain why CNN models arrive at their conclusions. Other then highlighting the patterns found in the interpretability study, we employ the extracted heatmpas to enhance Visual Transformers classifiers' performance and explanation quality. This approach yields substantial improvements in F1 score, ranging from 2\% to 8\%, across the datasets compared to benchmark values.
\end{abstract}


\begin{highlights}
    \item We conducted extensive replicability experiments across three malware datasets: Microsoft Windows BIG2015, Malimg, and a newly created dataset, VX-Zoo.
    \item Utilizing Class Activation Maps~(CAMs) provided valuable insights into the decision-making processes of the classifiers.
    \item The explainability of models was leveraged to enhance classifier performance through a novel image masking technique and to explain specific sample classifications.
    \item Our results demonstrated that HiResCAM outperforms GradCAM when applied with the masking technique.
\end{highlights}

\begin{keyword}
Malware \sep Explainability \sep Replicability \sep Neural Networks \sep Class Activation Maps \sep CNN
\PACS 0000 \sep 1111
\MSC 0000 \sep 1111
\end{keyword}

\end{frontmatter}


\section{Introduction}
\label{sec:intro}
With the evolution of malware threats, the Web has become a precarious space with the consequences of infections that extend beyond personal annoyance, profoundly resonating for businesses and governments. 
This situation has spurred a surge in cybersecurity investments, especially focused on robust anti-malware solutions. A fundamental task in the fight against computer threats is the categorization of malware in families, other than the simple detection. 
\par In recent years, besides traditional signature-based approaches, more contemporary methods rooted in machine learning and neural networks have been incorporated into the defender's toolkit, resulting in remarkable successes in countering malicious software~\cite{nataraj2011malware}\cite{naeem2019identification}. While traditional signature-based approaches were effective at classifying known malware, they struggled to keep up with modern threats' rapid evolution and polymorphic nature~\cite{limits}. However, machine learning algorithms, particularly neural networks, have revolutionized the field by enabling the proactive detection and mitigation of previously unseen and zero-day malware~\cite{ucci2019survey}. 


\par A widely used technique that harnesses the power of CNNs is visualization. Researchers can generate an image with the data extracted from a malicious file, and use it as a sample for the Convolutional Neural Networks (CNN)~\cite{nataraj2011malware}. This technique takes advantage of the convolutional neural network models' intrinsic qualities such as robust feature extraction, spatial invariance, adaptation, and efficient processing. 

One advantage is that CNNs almost remove the need for expert feature extractions, which can be provided only by a restricted number of specialists. The technique aims to minimize human involvement in routine tasks, thereby reducing the potential for error and allowing individuals to concentrate on more complex and nuanced aspects of their work. By automating repetitive processes, this approach not only enhances efficiency but also enables humans to leverage their unique cognitive abilities for tasks that require deep understanding and critical thinking~\cite{humanfactor1, ucci2019survey}. Another benefit is the fact that using neural networks on images makes the antivirus software more privacy aware, removing the need for access the whole raw binary. This process, applied to all new and potentially suspicious files, involves both static techniques (such as signature detection, PE file resource examination, and import/export table analysis) and dynamic methods (monitoring API calls, file operations, and registry modifications). Such comprehensive access to files risks information leakage to the antivirus program.
The visualization technique employed in our study is, in the worst case, as intrusive as the data used to generate the image. In high-privacy environments, the basic binary-to-image transformation could be performed by a trusted OS component, limiting the antivirus software's access to only the resulting image and thus reducing privacy risks. 
While privacy may seem peripheral to malware classification, recent years have seen governments request research into potential attacks by rogue antivirus vendors, underscoring its growing importance in cybersecurity discussions~\cite{privacy1}.
\par The absence of interpretability presents challenges, particularly in critical domains like healthcare, legal systems, or cybersecurity, where the need for explainability and transparency is paramount. The demand for explainable outcomes is steadily growing, especially among legislators who seek to ensure privacy, fairness and accountability in the decisions made by artificial intelligence systems. A prime example of this is the European Union's General Data Protection Regulation~(GDPR)~\cite{gdpr}, aimed at safeguarding individuals' privacy rights and governing the processing of their data. Similar laws are being deliberated in the United States~\cite{equal_credit_opportunity_act} and China~\cite{china_cybsec_law}. 

One of the significant drawbacks of neural networks is the inherent challenge of explaining the classification results to humans. Explainability is a critical issue in various fields, especially when the analyzed models are based on deep learning. While neural networks excel in complex pattern recognition tasks, their inner workings and decision-making processes can be challenging to interpret and understand. Unlike traditional machine learning algorithms used in visualization, such as decision trees, SVMs and in particular hidden markov models~(HMM), which provide explicit rules or feature importance rankings, neural networks lack explicit transparency~\cite{iadarola2021towards, hmm1, hmm2}. This lack of openness arises from the highly interconnected layers, nonlinearity and numerous parameters within the network. As a result, determining which specific features contribute to a particular classification decision can be elusive. Furthermore, the field has been traditionally dominated by the pursuit of performance metrics, with researchers primarily focused on improving accuracy rather than interpretability. An additional complexity emerges in the practical application of explainability studies: the interpretation of results is not straightforward, especially in domains like malware classification. Unlike standard image classification where artifacts are easily identifiable, detecting meaningful features in malware images requires domain-specific expertise, making the translation of explainability insights into actionable intelligence a significant problem. We addressed these challenges by systematically investigating resources from previous studies, meticulously implementing the CNN architectures using state-of-the-art libraries, rigorously following paper methodologies, and carefully tuning undocumented hyperparameters through extensive empirical experimentation. We finally tested the reproduced models on benchmark and new dataset to study their real comparable effectiveness. We developed a method to exploit CAM heatmaps by creating cumulative class-level heatmaps to identify characteristic macro areas for each malware class. This approach enables malware analysts to compare individual sample heatmaps against the cumulative heatmap, verifying artifact consistency in highlighted regions. By masking areas deemed less important and retraining classifiers on these modified images, we demonstrated improved classification performance, validating the CNN's focus on critical image regions.
The replicability of the results obtained by the best model is of prime importance~\cite{daoudi2021lessons}, especially in a crucial field like cybersecurity, where millions are spent to get the best results. In the context of malware analysis, where the complexity of threats and the sophistication of cyber-attacks are constantly evolving, replicability acts as a litmus test for the reliability of detection methods, classification algorithms, and mitigation strategies. Replicability has rarely been tackled by researchers due to its inherent challenges, primarily stemming from the lack of specific implementation details in published papers and the lack of codebases shared by authors. Researchers often omit critical technical specifics, making it difficult for others to precisely reproduce their experimental setups and results. Replicable research allows experts to scrutinize methodologies, identify potential biases, and validate the effectiveness of proposed solutions. Furthermore, it fosters a culture of collaboration and knowledge-sharing within the research community, enabling the collective development of robust and reliable tools to combat ever-changing malware threats.

\par With this study, we aim to tackle these two separated challenges to model robustness: \textit{replicability} and \textit{explainability}. Utilizing interpretability techniques on established models developed by reputable research groups lends greater relevance to the study. This approach not only validates the interpretability tools themselves, but also provides valuable insights into the inner workings of widely recognized models, enhancing the overall impact and applicability of the research. We tested replicability by selecting six of the best-performing CNNs found in the literature and implementing the architectures using the information found in the official documents presenting them. We analyzed the issues that emerged and studied the results, comparing them to the ones reported. After having produced realistic models, we moved on to the study of explainability. We employed a technique called Class Activation Maps~(CAM) to address the challenge of explainability. In particular, we are the first to implement High-Resolution Class Activation Map~(HiResCAM)~\cite{hirescam} for accurately explaining the state-of-the-art CNNs used in developing malware scanners. HiResCAM provides insights into the decision-making process of CNNs by highlighting the regions of an input, in our specific case an image, that contribute the most to a specific classification outcome. 

The article poses the following research questions:
\begin{itemize}
       
    \item \textbf{RQ1:} Can researchers reasonably replicate or reproduce the models showcased in a state-of-the-art paper? What are the current issues affecting the field of malware image visualization?
    
    \item \textbf{RQ2:} How can heatmaps help shed light on the functioning of CNNs in visualizing malware, offering insights into their mechanisms? 
    
    \item \textbf{RQ3:} Can these insights be extrapolated to evaluate qualitatively models and enhance other classifiers?

    \item \textbf{RQ4:} How does the inclusion of a new dataset impact the generalization capabilities of existing models?
    
\end{itemize}

The main contributions of this paper are the following:
\begin{enumerate}
\item To better gauge the performance of the analyzed models, we created a new dataset (VX-Zoo) with different characteristics from the other two usually chosen in the literature to validate our results further (Section~\ref{sec:413});
\item To study the robustness of state-of-the-art research models found in the literature, we carry out a replicability study on six models presented in five high-quality papers in the domain of malware image-based classification, highlighting the challenges that researchers have in the implementation of these models (Section~\ref{sec:51});
\item To investigate the explanation of the replicated models, we produce heatmaps using two different CAM methodologies to show how the individual neural networks classify the samples. In particular, we analyze the cumulative heatmaps to comprehend the way different neural networks arrive at their final prediction and the errors they might make (Section~\ref{sec:expl});
\item Finally, to exploit the explainability step in a way that can impact classifiers performance, we introduced a new and upcoming technique in malware visualization called Visual Transformer and show how the study of CNNs heatmaps can help improve classifiers results (Section~\ref{sec:53}).

\end{enumerate}

\section{Background}
In this section, we will explain the main ideas on which our research is based. First, we will talk about how people visualize malware and focus on the common deep learning models. We will also discuss the replicability framework needed to implement the latest models. From there, we investigate the explainability techniques used to study CNNs in other field and how researchers have started to apply them to malware image classifiers.
\subsection{Malware Visualization}
Malware visualization is a technique used in malware analysis to visually represent the behavior and characteristics of malware~\cite{gopinath2023acomprehensive}. 
Visualization involves creating graphical representations of various aspects of the malware, such as its network activity, file activity, system calls, or the simple binary file without additional enhancements. Malware developers often create new variants by making small changes to the code, allowing for the identification of related malware strains. Analysts can identify similarities and differences between different malware classes by representing the behavior and characteristics of malware in a visual format. 
This approach bypasses the requirement for disassembly, which is the non-trivial~\cite{disass} process of reading the binary and transforming it in assembly language or execution, which are common in existing classification methods, while still achieving significant performance improvements. Moreover, it proves to be robust against prevalent obfuscation tactics like section encryption~\cite{vasan2020imcec}. The pipeline used in malware classification through malware visualization comprises the following steps: information gathering, image generation, feature extraction, and classification. 
\begin{figure}[b!]
     \centering
     \begin{subfigure}[b]{0.24\linewidth}
         \centering
         \includegraphics[width=1\linewidth]{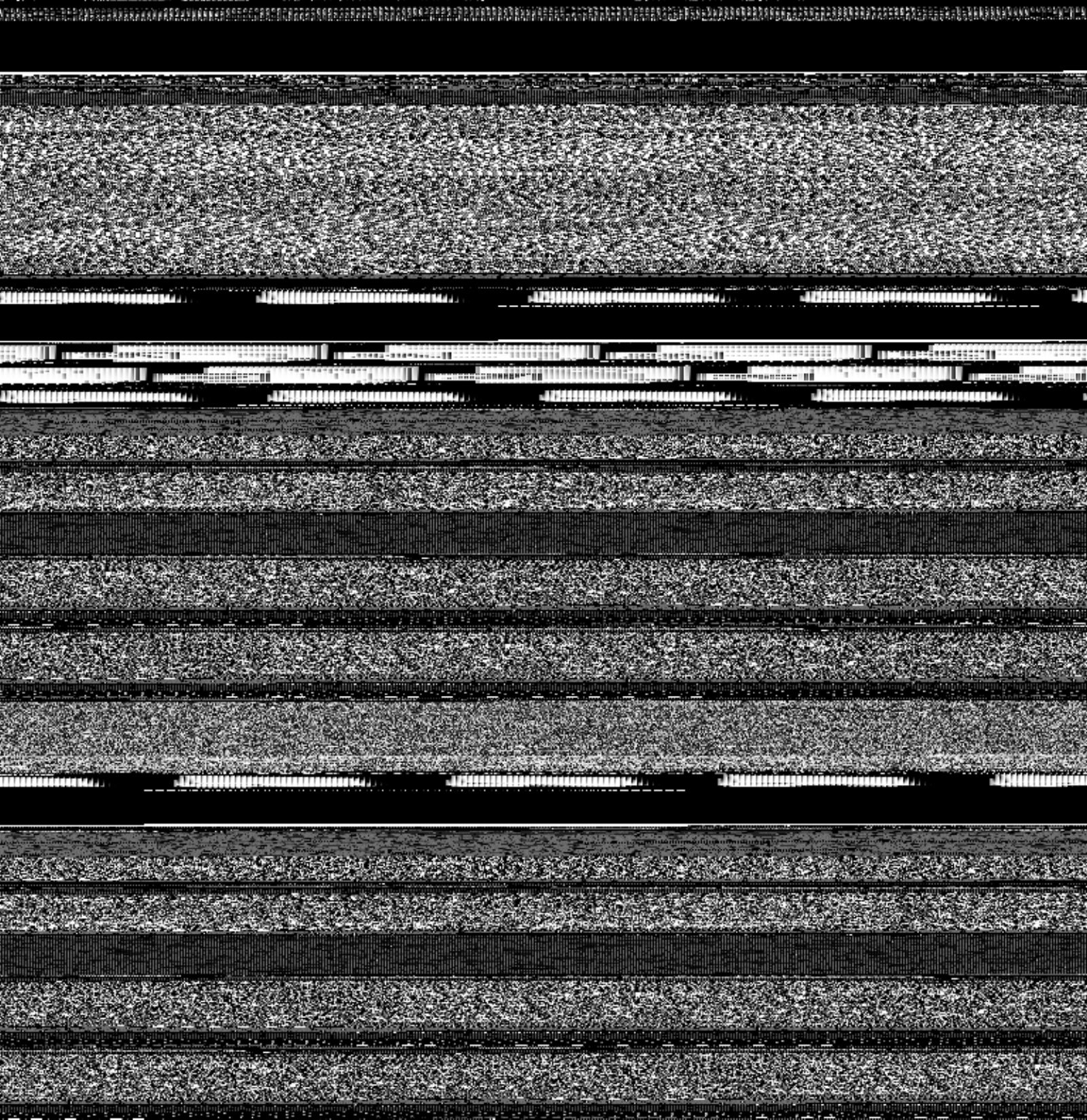}
         \caption{003f8...06716}
     \end{subfigure}
     \hspace{0.07\linewidth}
     \begin{subfigure}[b]{0.24\linewidth}
         \centering
         \includegraphics[width=\linewidth]{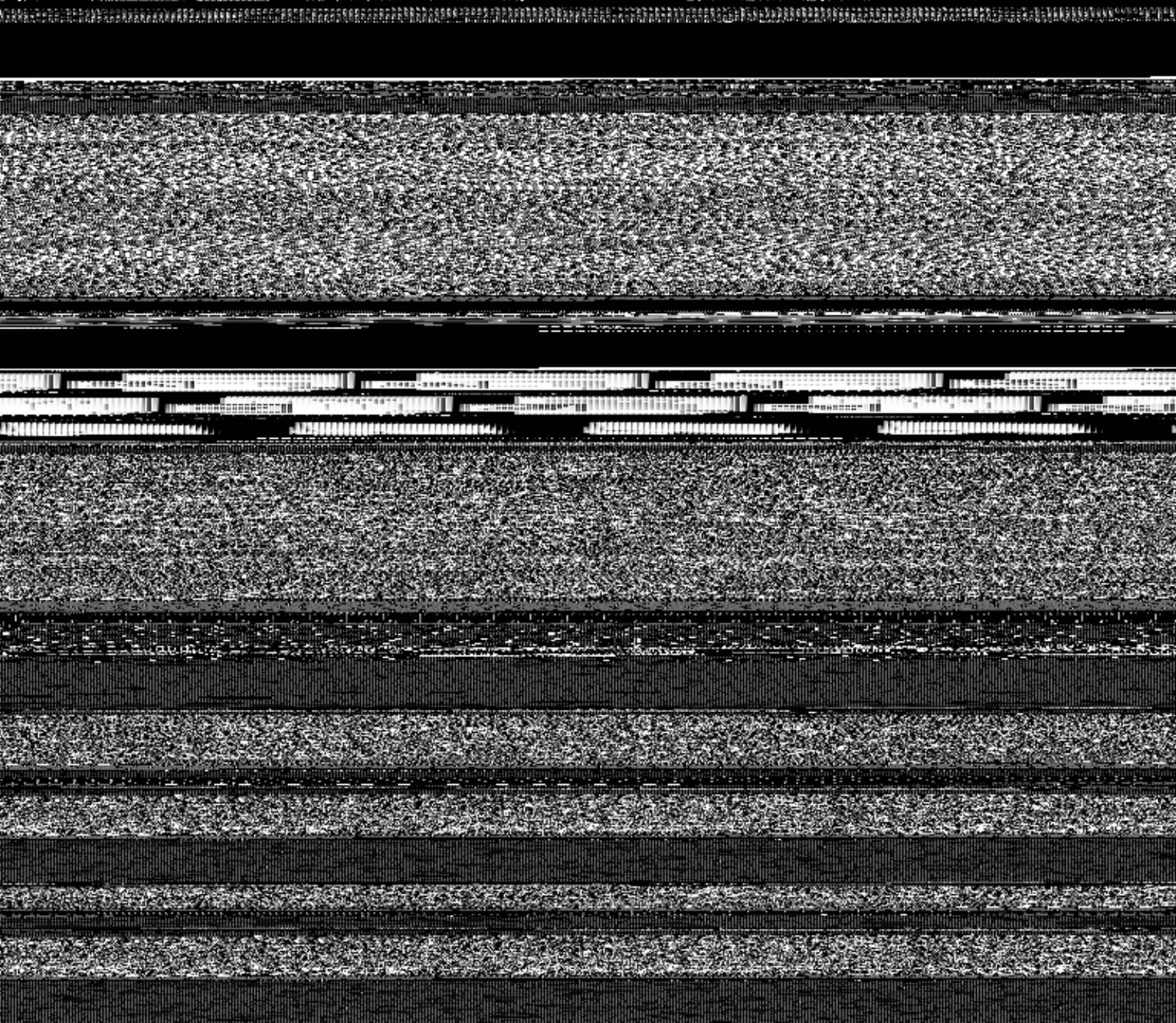}
         \caption{04a15...4f036}
     \end{subfigure}
     \hspace{0.07\linewidth}
     \begin{subfigure}[b]{0.24\linewidth}
         \centering
         \includegraphics[width=\linewidth]{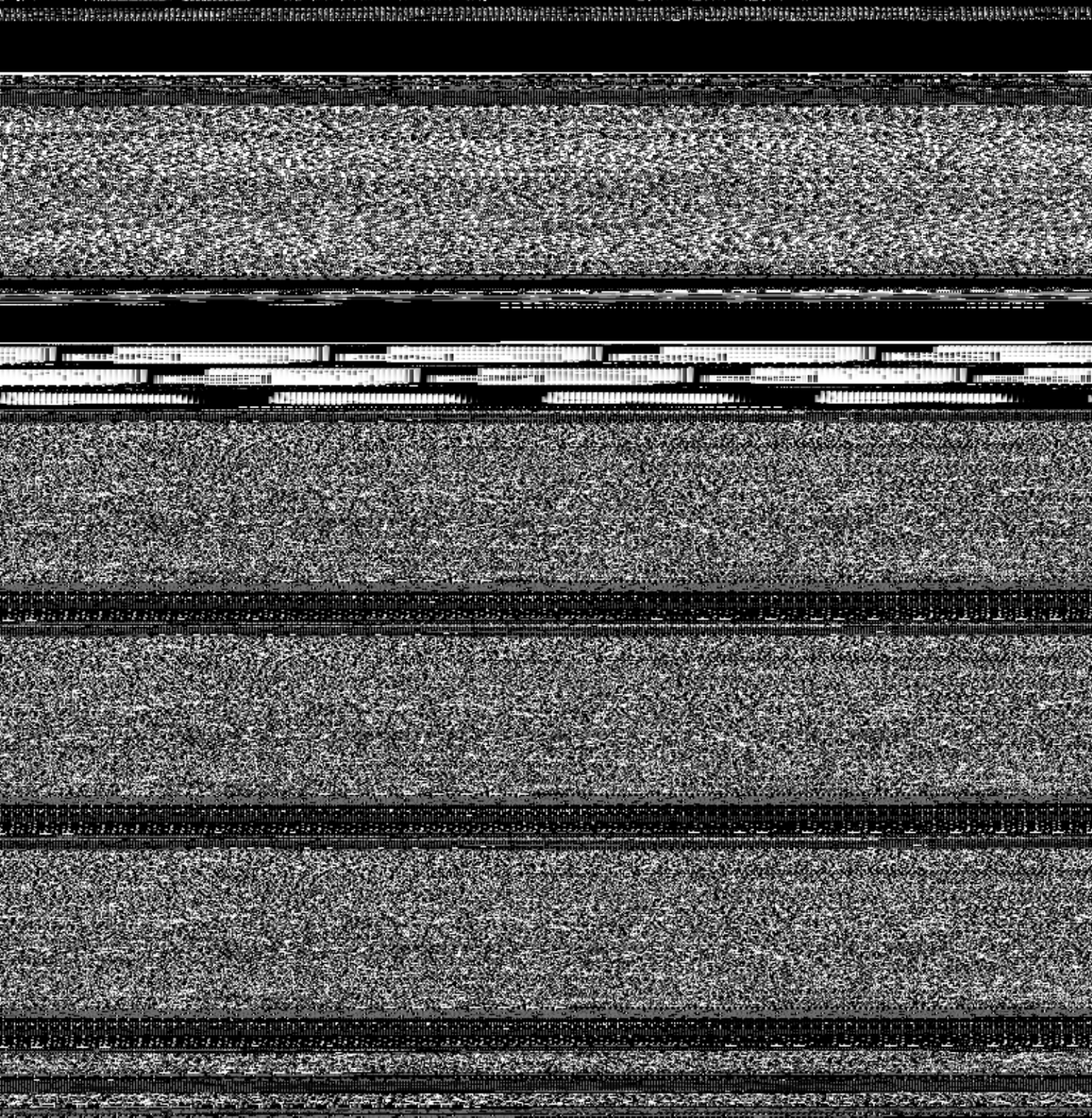}
         \caption{04a17...e7756}
     \end{subfigure}
     \caption{Example of grayscale images from the VB.AT MalImg dataset.}
        \label{fig:grayscale_example1}
\end{figure}
Researchers can employ different methods to transform malware into an image representation, with the first consideration being the information available to them. This information can range from the raw bytecode of the malware to a memory dump of the executed code from a sandbox~\cite{bozkir2021catch}. 
The standard methodology applied to transform a malware into a grayscale image is the so-called \textit{line-by-line technique}~\cite{nataraj2011malware}. 
Figure~\ref{fig:grayscale_example1} illustrate striking resemblances in the texture and arrangement of images representing malware from the same family. The next step in developing a reliable malware classifier is to extract relevant features from the generated image of the executable. 
The last step involves classifying the samples using the extracted features. At this stage, any machine learning or deep learning classifiers commonly used in other domains can be applied, as the information has been transformed and synthesized into a format suitable for input into a classifier. If using CNNs, these two last steps are usually combined and the researcher has just to design and train the model.

Few authors have asked themselves the question of replicability. The preliminary study present at the beginning of most papers showing comparable models does not usually include a discussion about implementing them but often reports only the results taken at face value. Most researchers' lack of sharing of the codebases further exacerbates this problem. It is essential to understand that the concept of replicability has varied over time and can mean different things to different people. Daoudi et al.~\cite{daoudi2021lessons} have helpfully defined three concepts to represent different ways researchers can implement a model with various degrees of faithfulness. These three concepts are \textit{repeatability}, \textit{reproducibility} and \textit{replicability}. The difference between those three lies in which team carries out the new experiment and which experimental setup is used. Researchers talk about repeatability if the same team performs the original experiment on the same experimental setup. Reproducibility, if another group reproduces the experiment on the same setup. If, in the end, another team experiments on a different experimental setup, we are talking about replicability. Reproducibility and replicability appears to be the key concern if the scientific community wants to validate the results obtained by a research group. The replicability of the results obtained by the best model is of paramount importance in the context of malware analysis, where the complexity of threats is constantly evolving. Replicable research allows experts to scrutinize methodologies, identify potential biases, and validate the effectiveness of proposed solutions. 

\subsection{Explainability}
The need for transparency and accountability in automated decision-making systems has made explainability a crucial element in machine learning. Explainability holds foremost importance in cybersecurity, where ambiguity is unacceptable. Although some machine learning algorithms inherently explain their outputs, widely used neural networks lack inherent explainability~\cite{iadarola2021towards}. This lack of transparency has raised concerns and led to a demand for interpretability techniques to enhance trust, mitigate bias, and ensure the ethical deployment of machine learning models. The purpose of any explanation generally falls into one of two categories: (i) providing insights into model training and generalization; these explanations offer practitioners valuable information for model training and validation decision-making, such as the quantity of labeled data, hyperparameter values, and model selection. (ii) explanations elucidating model predictions, helps practitioners understand why the model made a specific prediction, typically related to the input data. 
The explanations employed in our research specifically fit into this second category, facilitating the clear communication of model predictions to individuals lacking expertise in the subject matter. 
In the case of malware images, explanations are in the form of saliency maps or heatmaps. Both highlight the crucial regions that influenced the network's prediction. It's important to note that while heatmaps are the prevalent method for presenting model explanations, their interpretation is subjective and dependent on the practitioner~\cite{xaisurvey}. The addition of an explainability study over a simple implementation of a model might be seen as cumbersome, but in the situations in which reasonableness and transparency are necessary, it becomes mandatory. Legislators have also identified the need for accountability in the decisions made by artificial intelligence systems in documents like the GDPR~\cite{gdpr} and its right to explain provision. The use of explained classifiers, like the ones developed in this study, can be the starting point for a comprehensive, interpretable malware classification system.

\section{Related work}
\label{sec:related}
In this section, we provide a brief overview of literature relevant for our research. Each referenced work is summarized to highlight its key contributions and relevance to the field. 
\subsection{Machine Learning}
The introduction of machine learning techniques has been proven fruitful in various domains and cybersecurity use is increasing~\cite{ucci2019survey}. In 2011 Nataraj et al.~\cite{nataraj2011malware} proposed an algorithm to transform the binary file of malicious samples in grayscale images. GIST texture features have been extracted from the malware image dataset to feed an SVM classifier. The model's accuracy was tested on the newly introduced MalImg dataset, achieving an accuracy rate of 97.18\% .

\par In 2020, Roseline et al.~\cite{roseline2020intelligent} proposed an anti-malware system that employs a layered ensemble of random forests, mimicking deep learning techniques. This approach does not require hyperparameter and operates with reduced model complexity. The system proposed demonstrated detection rates of 98.65\% and 97.2\% for Malimg and the other standard benchmark dataset Big2015~\cite{ronen2018microsoft}, surpassing other cutting-edge methods of the time.

\subsection{Deep Learning}
The utility of machine learning techniques has been overshadowed by deep learning methods' increasingly prevalent and superior performance. Furthermore, the requirement for accurate feature extraction, often necessitating expert assistance, represents another vulnerability of machine learning approaches that is not present with neural networks.

\par In 2019, Gibert et al.~\cite{gibert2019using} employ the visualization technique by implementing a custom CNN. This neural network handles feature extraction independently, eliminating the necessity for additional algorithms. The model exhibits a 1.30\% accuracy boost over the initial Nataraj et al. model, showcasing the effectiveness of CNNs.
\par In 2019, authors in Beppler et al.~\cite{beppler2019laying} focused on texture analysis for malware classification using in-the-wild samples, critically examining the application of this approach for detecting and categorizing malicious software, testing multiple scales, descriptors, and classifiers. By reproducing literature experiments, they discovered that texture analysis may be unfeasible when using existing malware representations, and identified significant biases in sample selection that could produce unrealistic results. Their results in this sense are similar to what we observed in our replicability study. Through testing with a broader, unfiltered dataset, the authors ultimately concluded that texture analysis presents substantial challenges for accurate malware classification, especially when confronting sophisticated obfuscation techniques across diverse malware families.
\par In 2020, Vasan et al.~\cite{vasan2020imcfn} experimented with a simplified version of the \textit{VGG16} CNN. The researchers employed RGB images by applying a basic colormap to the originals. They then utilized transfer learning alongside fine-tuning to achieve a competitive model. In this study, only the dense and initial convolutional layers underwent training, resulting in a final accuracy of 98.82\% on the MalImg dataset. In the same year, the same authors~\cite{vasan2020imcec}, used an ensemble technique to harness the power of different feature extractors in the form of different CNNs and other classifiers. The fine-tuned models employed in the study have been \textit{VGG16} and \textit{ResNet50}. The final accuracy result on the MalImg dataset was 99.50\%, marking a big improvement over previous models and, most importantly, without the need for complicated image generation procedures.  
\par In the context of explainability, in 2021, the researchers in~\cite{iadarola2021towards} introduced an interpretable method for identifying malware in the Android ecosystem. They utilized the GradCAM algorithm to enable visual debugging of models, providing insights into the particular areas that play a crucial role in predicting the malicious nature of images associated with Android applications.
\par In~\cite{tekerek2022anovel}, the authors propose a method utilizing \textit{DenseNet} CNNs, grayscale and RGB images. They standardize the technique for file transformation in grayscale images and use a CycleGAN-based data augmentation method to handle imbalanced data sizes among malware families. Testing on Big2015 and DumpWare10 datasets showed significant improvements in classification accuracy, reaching 99.86\% for Big2015 and 99.60\% for DumpWare10, demonstrating the effectiveness of the proposed method.
\par In~\cite{shaukat2023anovel}, the authors introduce a novel approach to malware detection by combining static and dynamic analysis techniques. The proposed method visualizes PE files as colored images, extracts deep features using a fine-tuned deep learning model. Experimental validation involving 12 machine learning and 15 deep learning models, conducted on various benchmark datasets, shows an accuracy of 99.06\% on the Malimg dataset and demonstrated statistical significance through rigorous testing methods. 
\par In~\cite{mctvd}, Deng et al. propose a malware classification method based on three channel visualization called MCTVD, which employs small, uniform-sized malware images and a shallow convolutional neural network. Experimental results demonstrate MCTVD's effectiveness, achieving an accuracy of 99.44\% on Big2015 under 10-fold cross-validation.
\par In 2023, Nguyen et al.~\cite{gan1} introduced a technique to test the robustness of malware visual classifiers by employing a GAN and different feature extraction methods used for the images. Their study is important because it shows an interesting adversarial attack in which fake malware images are generated and fed to the classifier. They also demonstrate that CNNs, specifically \textit{ResNet} and \textit{VGG}, remain the most advanced classifiers in the domain of malware visualization. 
A comparable conclusion has been reached by Prajapati et al.~\cite{stampbook}. The authors conducted a comprehensive investigation on a novel dataset comprising diverse classifiers utilized in the field, with particular emphasis on CNNs, LSTM, MLP, and GRU. The final results comparison shows how the same \textit{ResNet} and \textit{VGG} CNN families outperform other classifiers, providing the most promising classifiers' architecture for the years to come.

\par Deep learning, especially CNNs, overwhelmingly dominates the current landscape of malware detection using image-based methods. Researchers extensively validate these architectures, either borrowed from other disciplines or simpler ad-hoc models, consistently achieving cutting-edge performance. To enhance their models, researchers typically pursued two avenues: refining classification using superior classifiers, primarily CNNs, or improving image generation methodologies to create richer, more informative images.
Ongoing refinement of CNNs, often through ensemble techniques, emerges as the favored approach for achieving optimal outcomes. This shift from machine learning to CNNs undeniably bolsters malware categorization's effectiveness but also introduces a ``black-box" approach, obscuring the inherent transparency characteristic of traditional machine learning models~\cite{iadarola2021towards}. Few researchers have delved into the intricacies of why specific choices are made, such as selecting the best-performing neural networks for ensembling. This lack of clarity often leads to situations where combining the top-performing neural networks does not consistently generate the best final results. The field, therefore, grapples with the trade-off between superior performance and the understanding of the decision-making processes underlying these advanced models. Table~\ref{classification_table} contains a summary of the most interesting papers.

\begin{table}

\scriptsize

\begin{tabular}{|l|c|c|c|c|c|}
\multicolumn{1}{c}{\rotatebox{90}{\textbf{Paper}}} & \multicolumn{1}{c}{\rotatebox{90}{\textbf{Year}}}  & \multicolumn{1}{c}{\rotatebox{90}{\textbf{Image type}}} & \multicolumn{1}{c}{\rotatebox{90}{\textbf{FE}}} & \multicolumn{1}{c}{\rotatebox{90}{\textbf{Classifier}}} &\multicolumn{1}{c}{\rotatebox{90}{\textbf{Interpretability}}}\\ \hline
Nataraj et al.~\cite{nataraj2011malware} & 2011 &Grayscale &GIST &  KNN & \\ \hline
Gibert et al.~\cite{gibert2019using} & 2019 & Grayscale &  & CNN &  \\ \hline
Vasan et al.~\cite{vasan2020imcec} & 2020 & Grayscale & & CNN, SVM & \\ \hline
Iadarola et al.~\cite{iadarola2021towards} & 2021 & Grayscale& & CNN & \Checkmark \\ \hline
Tekerek et al.~\cite{tekerek2022anovel} & 2022 & Grayscale, Color && CNN &\\ \hline
Shaukat et al.~\cite{shaukat2023anovel} & 2023 & Color && CNN, SVM &\\ \hline
Deng et al.~\cite{mctvd} & 2023 & SFC, Markov, & & CNN &\\ 
 &  & Hash, Color & & &\\\hline

\end{tabular}

 \caption{Summary of the state-of-the-art analyzed papers.\label{classification_table}}
\end{table}

\newcommand\Tstrut{\rule{0pt}{2.6ex}}       
\newcommand\Bstrut{\rule[-0.9ex]{0pt}{0pt}} 
\newcommand{\TBstrut}{\Tstrut\Bstrut} 

\section{Proposed Methodology}
\label{sec:methods}
In this section, we delve into the methodology employed to analyze and address the research questions posed in the introduction. Our exploration starts with a comprehensive overview of the three datasets, detailing the methods employed for data acquisition and emphasizing the nuances that distinguish them. We explain the rationale underpinning our selection of six CNNs derived from five researched papers of high reputation. Transitioning from replicability considerations, we introduce the array of explainability tools utilized, predominantly focusing on CAMs, specifically High Resolution CAM~(HiResCAM)~\cite{hirescam} and GradCAM~\cite{gradcam}. Our exploration navigates through their unique implementations and distinctive features.
Furthermore, we introduce the Structural Similarity Index~(SSIM)~\cite{ssim} metric and articulate its significance within the context of our research. We present the cumulative heatmap for a particular malware family and cumulative-SSIM, a new metrics introduced in this research. This metric will be used to identify the best CNN in terms of explainability. Toward the section's end, we clarify our rationale for adopting a Vision Transformer~(ViT)~\cite{vit} as an additional testing classifier for our masking method. 

\subsection{Datasets}
The prevalent datasets utilized in existing literature, namely MalImg~\cite{nataraj2011malware} and Big2015~\cite{ronen2018microsoft}, stand as the de facto benchmarks. Opting to utilize both, we aim to yield substantial findings that validate the robustness of our research. Creating a novel dataset, VX-Zoo, specifically designed for this study, adds an original dimension to our exploration. VX-Zoo, albeit maintaining the same structure of Big2015 and MalImg, contains more samples than both. Detailed presentations of these three datasets follow in subsequent sections.
\begin{table*}[ht]
\small
    \centering
    \begin{tabular}{|l|l|l|l|l|l|l|l|l|l|l|}
    \hline
    \textbf{Family} & \textbf{Total}&\textbf{Packed} &\textbf{2015} & \textbf{2016}&\textbf{2017} &\textbf{2018} &\textbf{2019} &\textbf{2020} &\textbf{2021} &\textbf{2022} \\ \hline
    AntiFW         & 5777 &99\% & 2784 & 1419 & 1574 & 0 & 0 & 0 & 0 & 0     \\ \hline
    Expiro         & 364 &79\% & 325 & 0 & 0 &0 &0 & 0& 1 &      38  \\ \hline
    Lamer         & 754 &100\%  & 16 & 0 & 0 & 17 & 27 & 636 & 44 & 14  \\ \hline
    Parite          & 380& 99\%  & 361 & 7 & 1 & 1 & 7  &1 &2 &0     \\ \hline
    Sality          &  855& 89\% & 809 & 7 & 1 &10 &15 &11 &2 &0     \\ \hline
    Virut          & 1612 &84\% & 849 & 11 & 5 &7 &15 &647 &39 &39  \\ \hline
    WBNA        & 908 &5\% & 885 & 0 & 0 &2 &1 &1 &12 &7    \\ \hline
    Wabot          & 442 &100\%  & 60 & 0 & 0 &0 &0 &363 &4 &  15  \\ \hline
    Vtflooder          & 465 &100\%& 465 & 0  & 0&0 &0 &0 &0 &0    \\ \hline
    Alman          & 226 &45\% & 218 & 3 & 0 &0 &2 &2 &1 &0   \\ \hline
    Elkern          & 219 &100\% & 219 & 0 & 0 & 0 & 0& 0& 0&0  \\ \hline
    Nimul          & 3044 &99\%  & 37  & 0 & 0 & 4& 2& 7& 2984 & 10  \\ \hline
    Debris          & 208 &56\%  & 208 & 0 & 0 & 0 & 0 & 0 & 0 & 0 \\ \hline
    Inject     & 275 &100\% & 0 & 0 & 275 & 0& 0&0 &0 &0    \\ \hline
    VB        & 130 &0\% & 0 & 0 & 0 &0 & 0& 120&4 & 6 \\ \hline
    \end{tabular}
        \caption{VX-Zoo dataset composition divided by number of samples found for each year. The total percentage of packed samples for each class is provided in the second column. \label{tab:table-ogds}}

\end{table*}
\subsubsection{MalImg and Big2015}
MalImg stands out as the pioneering malware dataset exclusively designed for image visualization. Despite its age, it remains a cornerstone in the realm of Windows malware image-based analysis, and it continues to be widely utilized. Introduced by Nataraj et al. in their seminal paper~\cite{nataraj2011malware} in 2011, this dataset sparked the interest in malware visual classification. Comprising 9,339 malware grayscale images categorized into 25 distinct classes, this dataset is publicly accessible. 
\par The Microsoft Malware Classification Challenge Dataset, often called Big2015 or MMCC~\cite{ronen2018microsoft}, is the second most cited repository in the field. This dataset encompasses a collection of 10,860 malware executables, categorized into 9 distinct families. Originating from the Microsoft Malware Classification Challenge in 2015, participants were tasked with training their classifiers, of varied types, on these 10,860 malware samples. The challenge required participants to predict on a separate set, the results of which were evaluated remotely.

\subsubsection{VX-Zoo}
\label{sec:413}
While Big2015 and MalImg possess unique strengths and weaknesses, we recognized the necessity of generating a new and more contemporary malware set to establish additional benchmarks. We introduce the fundamental principles that guided the development of this new dataset and the differences in characteristics between the benchmark datasets.

\begin{enumerate}
    \item \textbf{Incorporating new malware}: Acknowledging the evolving malware landscape, we aimed to create a dataset that reflects the present-day scenario. MalImg and Big2015 have started showing signs of aging, and testing on them may not provide a realistic perspective on modern malware variants;
    
    \item \textbf{Year-wise identification}: Recognizing the significance of analyzing malware from the same class but across different years, we tried to include malware samples from diverse years whenever possible. This approach, which is novel compared to other malware datasets, enables researchers to assess classifier performance when confronted with malware that has undergone mutations over time. We considered malware families that are not necessarily new, but that appear consistently in modern times. Malware is continually updated, and code from older families can also be reused in more modern variants that are traceable to old classes. A modern AV has to be able to detect older samples as much as new ones, especially if the older families appear frequently even in modern times. For this reason older families whose signatures are still being updated and studied by AV detectors~\cite{virut, sality, expiro, nimnul};
    
    \item \textbf{Maintaining imbalanced characteristics}: Emulating the natural variation observed in the wild, where malware classes are not evenly distributed, we preserved the imbalanced nature of sample distribution. This decision aligns with the reality of cybersecurity threats, where certain malware classes are widespread, while others are more niche;
    
    
    \item \textbf{No distinction between obfuscated and non-obfuscated malware}: Unlike some datasets, among which MalImg, that categorize all obfuscated malware into a single category, we opted not to distinguish between obfuscated and non-obfuscated malware. Our objective was for the classifiers to correctly identify malware classes, regardless of whether the malware was obfuscated or not. For this reason we found ourselves with classes, like \textit{Elkern}, \textit{Inject} and \textit{Vtflooder} for which samples where always packed (using UPX), while in other classes only a subset of the samples was found packed in some way. This situation, and the different proportions of packed / unpacked malware, found in the wild have been maintained in VX-Zoo;
    
    \item \textbf{Increased sample size and intermediate class count}: In our endeavor to enhance the dataset’s robustness, we aimed for a larger sample size than both Big2015 and MalImg. Additionally, we settled on a number of classes that fell between the extremes, striking a balance that allowed for a comprehensive evaluation of classifier performance.
\end{enumerate}

\par To compile our samples, we accessed the malware binaries from VirusShare\footnote{\href{https://virusshare.com}{VirusShare}, Accessed: 2023-5-15}, stored within Zip files available through torrents. Spanning a period of eight years from 2015 to 2022. We randomly chose two Zip files uploaded during a specific year for each year, utilizing them as the foundational data source for that particular time frame. The labeling system utilized was based on the Kaspersky\footnote{\href{https://www.kaspersky.it/}{Kaspersky}, Accessed: 2023-5-15} nomenclature, obtained from the VX-Underground\footnote{\href{https://www.vx-underground.org/}{VX-Underground}, Accessed: 2023-5-15} repository. Leveraging these labels, we could discern and filter out non-PE~(Portable Executable) malware files present within the zipped archives. Subsequently, we focused in on 15 malware families that exhibited consistency across the eight years under consideration. Specifically, we focused on families that appeared at least three times, except for 4 classes (\textit{Inject.ahqtx}, \textit{Debris.b}, \textit{Elkern.b} and \textit{Vtflooder.ekl}). The resulting dataset showcased the diverse characteristics of malware families over time. Some classes exhibited sporadic spikes in specific years, while others displayed consistent patterns or concentrated occurrences in consecutive years. This dynamic evolution of malware families was cataloged and is detailed in Table~\ref{tab:table-ogds} for a comprehensive reference.

\subsection{Replicated CNN Models}
\begin{table*}[h!]
\small
    \centering
    \begin{tabular}{|l|l|l|l|l|l|l|}
    \hline
    \textbf{CNN} & \textbf{Dropout}  & \textbf{LR} & \textbf{Optim} & \textbf{M} & \textbf{WD} & \textbf{Split}  \\ \hline
   VGG16\cite{vasan2020imcec}           & 0.2      &    5x10e-6 &  SGD & 0.9 & 5x10e-6 & \textbf{63/07}/30 \\ \hline
    ResNet50\cite{vasan2020imcec}           &0.2         & 5x10e-6  & SGD & 0.9  & 5x10e-6 & \textbf{63/07}/30 \\ \hline
    IMCFN\cite{vasan2020imcfn}         & \textbf{0.2}          & 5x10e-6 & SGD & \textbf{0.9} & 5x10e-6 & \textbf{63/07}/30 \\ \hline
Gibert\cite{gibert2019using}          & \textbf{0.3}         & \textbf{5x10e-6}  & \textbf{Adam} & \textbf{0.9}  & \textbf{2x10-e3} & 10f\textbf{/100} \\ \hline
EfficientNetB0\cite{pratama2022malw}           & \textbf{0.0} & 1x10e-4 & Adam & \textbf{0.9} & \textbf{2x10e-6} & 68/12/20 \\ \hline
DenseNet121\cite{tekerek2022anovel}           & 0.3,        & 1x10e-3  & SGD & 0.9 & 1x10e-6 & 10f/20 \\ 
 & 0.2        &  &  &  &&\\ \hline
    \end{tabular}
    \caption{Hyperparameters of the analyzed CNN regarding the training of the neural network. Values in bold were not explicitly indicated and have been tuned or deduced from the paper content. [f] indicates the use of k-fold cross validation}
    
    \label{tab:table-hyperparams2}
\end{table*}
The issue of replicability in science is particularly crucial in the field of malware analysis and visualization. This domain faces unique challenges that make reproducible research even more critical. Many authors claim state-of-the-art results while overlooking better-performing models, potentially skewing the field's understanding of progress. Given the highly competitive nature of malware detection, where improvements often range within 2–3\%, the ability to replicate results becomes paramount for validating claims and advancing the field. Furthermore, the lack of uniformity in data sharing practices among researchers hinders reproducibility efforts. Establishing standard protocols for sharing datasets, model architectures, and training parameters is essential. The complex, interdisciplinary nature of malware visualization, combining aspects of cybersecurity, data science, and visual analytics, further compounds these challenges. Addressing these replicability issues is crucial not only for scientific integrity but also for developing more robust and reliable malware detection and analysis techniques, which have significant real-world implications for cybersecurity.
We focused on using a variety of approaches instead of just picking the models that performed the best to study how well any type of malware classification CNN can be recreated reliably. Other than obvious characteristics like high quality publishing in a journal or conference and state-of-the-art results, we focused on CNN with particular quality. First, we only considered models tested on the known benchmark datasets Big2015 and MalImg to have comparable models. We then made an effort to vary the baseline architecture family (like \textit{VGG}, \textit{ResNet} and so on) and the year of publication. Ultimately, we prioritized the inclusion of models that employ transfer learning and fine-tuning as they represent superior approaches, at the time of writing, to obtain better models without incurring in significant expenditure on training.

The final six models chosen are:
\begin{enumerate}
    \item \textbf{Gibert CNN}: from Gibert et al.~\cite{gibert2019using}. A custom CNN architecture developed specifically for malware analysis;
    \item \textbf{VGG16}: implementation of the \textit{VGG16} architecture described by Vasan et al.~\cite{vasan2020imcec} as part of their ensamble architecture. Uses full model fine-tuning;
    \item \textbf{ResNet50}: implementation of the \textit{ResNet50} architecture described by Vasan et al.~\cite{vasan2020imcec} as part of their ensamble architecture. Uses full model fine-tuning;
    \item \textbf{IMCFN}: modified version of a \textit{VGG16} architecture provided by Vasan et al.~\cite{vasan2020imcfn}. Uses only last layers fine-tuning;
    \item \textbf{DenseNet121}: implementation of the \textit{DenseNet121} architecture provided by Tekerek et al.~\cite{tekerek2022anovel}. We ignored the CycleGAN technique that is added in the original paper to the model in the final test stage;
    \item \textbf{EfficientNetB0}: On of the \textit{EfficientNet} models implemented by Pratama et al.~\cite{pratama2022malw}. We selected the \textit{B0} version because it uses comparable size images as input. 
\end{enumerate}

\subsection{Models Interpretation}
\par In the context of explaining neural network models applied to image analysis, Class Activation Maps~(CAMs)~\cite{cam}, emerge as an invaluable technique. CAMs facilitate the discernment within an image of crucial regions instrumental in predicting a specific class. Typically utilized for object localization, saliency maps pinpoint the location of an identified object within an image. Our objective diverges slightly, as we aim to comprehend the areas considered pivotal by CNNs during the prediction process rather than identifying specific objects within bitmap images. The rationale behind this approach is identifying recurring image patterns prevalent across all malware samples within a particular family. This understanding aids in discriminating one class from another.
In this paper, we used two variants of CAM, GradCAM and HiResCAM with a particular focus on HiResCAM. In the next section we explain the differences with pros and cons of the two techniques. 
\begin{figure*}
     \centering
     \begin{subfigure}[b]{0.15\textwidth}
         \centering
    \includegraphics[width=\linewidth]{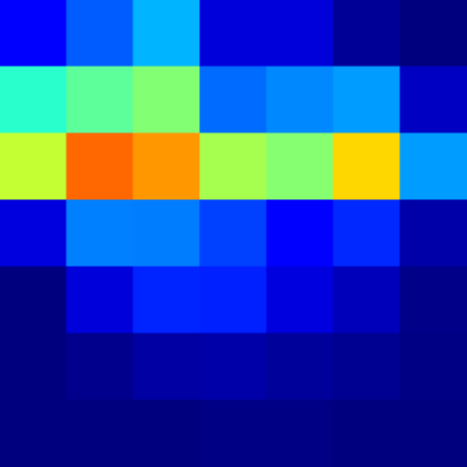}
    \caption{Agent.FYI} 
     \end{subfigure}
     \hfill
     \begin{subfigure}[b]{0.15\textwidth}
         \centering
    \includegraphics[width=\linewidth]{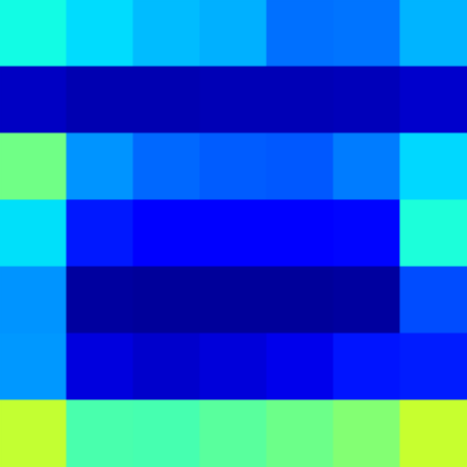}
    \caption{Khelios\_ver3} 
     \end{subfigure}
     \hfill
     \begin{subfigure}[b]{0.15\textwidth}
         \centering
    \includegraphics[width=\linewidth]{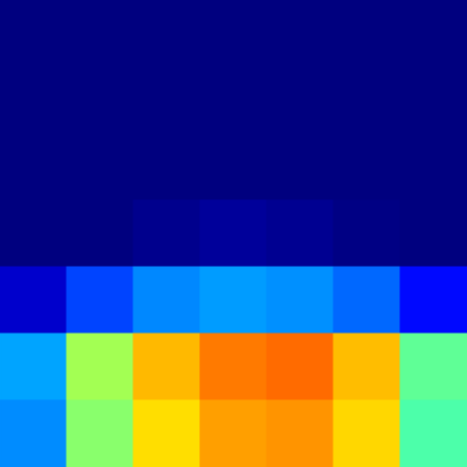}
    \caption{Rbot!gen} 
     \end{subfigure}
     \hfill
     \begin{subfigure}[b]{0.15\textwidth}
         \centering
    \includegraphics[width=\linewidth]{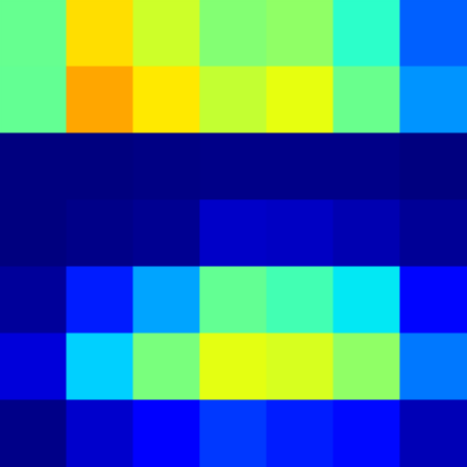}
    \caption{Skintrim.N} 
     \end{subfigure}
     \hfill
     \begin{subfigure}[b]{0.15\textwidth}
         \centering
    \includegraphics[width=\linewidth]{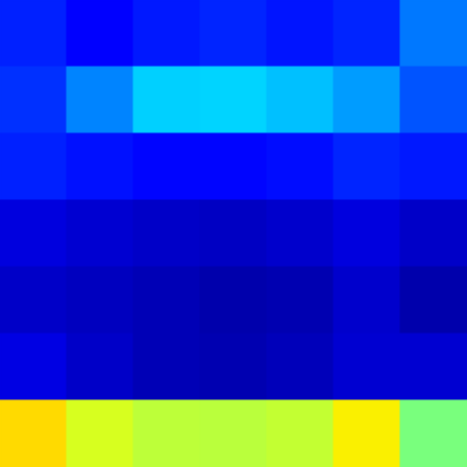}
    \caption{Wabot.a} 
     \end{subfigure}
    \caption{Example of cumulative GradCAM heatmaps extracted from malware families with different CNNs.} 
    \label{fig:gradcam_heatmaps}
\end{figure*}

\subsubsection{GradCAM}
Gradient-weighted Class Activation Mapping~(GradCAM)~\cite{gradcam} is an extension of the basic CAM technique. GradCAM highlights the important regions and provides a more localized and detailed visualization compared to CAM. Unlike CAM, GradCAM incorporates gradient information to weight the importance of the feature maps. It computes the gradients $\frac{\partial s_m}{\partial A_{d_1\,d_2}^f}$ of the predicted class score $s_m$ with respect to the activations in the final convolutional layer to identify important regions of an image. These gradients represent how sensitive the predicted class score is to changes in the feature map activations. It then calculates the importance of each feature map by taking the global average pooling~(GAP) of the gradients as shown in Equation~\ref{grad-1} 

\begin{equation}
\label{grad-1}
    \alpha_m^f = \frac{1}{D_1\,D_2}\:\sum_{d_1=1}^{D_1}\,\sum_{d_2=1}^{D_2}\,\frac{\partial s_m}{\partial A_{d_1\,d_2}^f}
\end{equation}

and indicated as $\alpha_m^f$ where $D_{1}$ and $D_{2}$ represent the height and width of the feature map. This provides a weight for each feature map $\textbf{A}^f$ indicating its contribution to the final prediction. The importance-weighted feature maps are then added to create the GradCAM heatmap, as shown in Equation~\ref{grad-2}
\begin{equation}
\label{grad-2}
 \Tilde{\mathcal{A}}_m^{GradCAM} = \:\sum_{f=1}^{F}\,\alpha_m^f \textbf{A}^f
\end{equation}

In Figure~\ref{fig:gradcam_heatmaps}, we display cumulative GradCAM images of example malware families.

\par Comparing with CAM, GradCAM offer the following improvements in image explanation (a) \textit{Incorporation of Gradient Information}: Gradient-based approach allows GradCAM to capture fine-grained details and more precisely localizes essential regions, (b) \textit{Sharper Object Boundaries}: GradCAM can provide detailed visualizations of object edges and parts, making it particularly useful for tasks like segmentation, and (c) \textit{Reduction of Overemphasis on Discriminative Parts}: CAM tends to highlight large discriminative parts, which might include irrelevant background regions. GradCAM, by utilizing gradients, reduces the tendency to overemphasize these areas, leading to more focused and relevant visualizations.

\subsubsection{HiResCAM}
 Draelos et. al~\cite{hirescam} investigated the differences of GradCAM and HiResCAM and proved that HiResCAM is a more general version of GradCAM. HiResCAM, like GradCAM provides detailed attention information over the spatial dimensions of the input data, allowing for precise localization of important regions used by the model for predictions. The concept originates from GradCAM's global average pooling phase, which can occasionally cause the attention maps to become blurred, reducing their capacity to highlight precise image areas that influence predictions. HiResCAM addresses the key limitation of GradCAM in how it handles the weights of feature map importance and how they interact with the components of the feature map. In GradCAM, each component of the final explanation must match the relative magnitudes and patterns of the feature map. This constraint leads to blurred visualizations as individual changes in the feature map (like rescaling or sign changes) are averaged out. HiResCAM preserves rescaling and sign changes in the individual elements of the feature map, ensuring that the high-resolution attention maps accurately reflect the model's computations. By maintaining these fine-grained details, HiResCAM provides a visualization that is not only high-resolution but also faithfully represents the model's decision-making process at the level of individual features. HiResCAM's ability to preserve fine-grained information is crucial for tasks where a detailed understanding of the model's reasoning is necessary. Practically, the formulas used by HiResCAM to calculate the attention map is Equation~\ref{hires-1}

\begin{equation}
\label{hires-1}
 \Tilde{\mathcal{A}}_m^{HiResCAM} = \:\sum_{f=1}^{F}\,\frac{\partial s_m}{\partial\textbf{A}^f} \odot \textbf{A}^f
\end{equation}

\begin{figure*}
     \centering
     \begin{subfigure}[b]{0.15\textwidth}
         \centering
    \includegraphics[width=\linewidth]{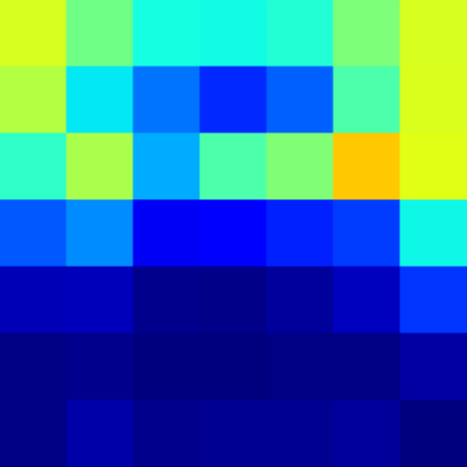}
    \caption{Agent.FYI} 
     \end{subfigure}
     \hfill
     \begin{subfigure}[b]{0.15\textwidth}
         \centering
    \includegraphics[width=\linewidth]{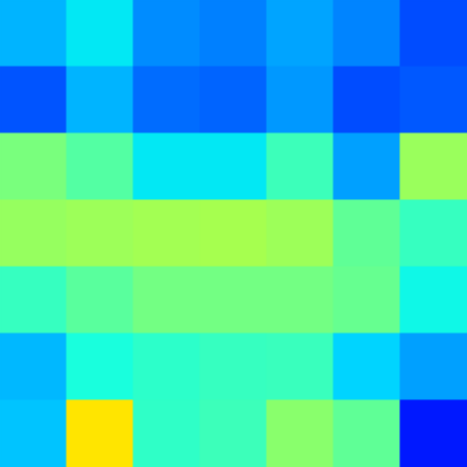}
    \caption{Khelios\_ver3} 
     \end{subfigure}
     \hfill
     \begin{subfigure}[b]{0.15\textwidth}
         \centering
    \includegraphics[width=\linewidth]{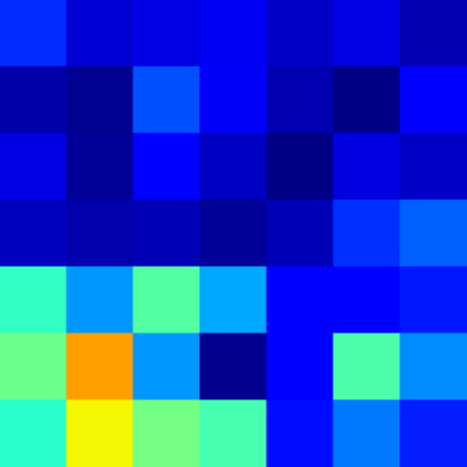}
    \caption{Rbot!gen} 
     \end{subfigure}
     \hfill
     \begin{subfigure}[b]{0.15\textwidth}
         \centering
    \includegraphics[width=\linewidth]{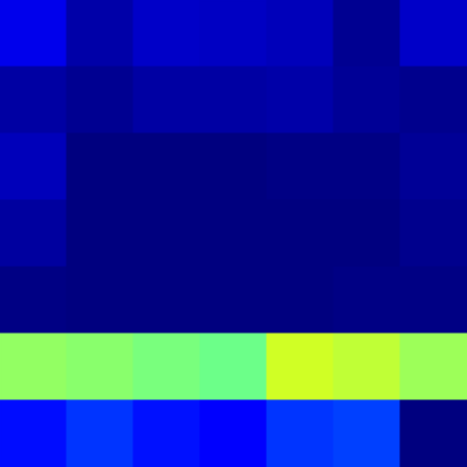}
    \caption{Skintrim.N} 
     \end{subfigure}
     \hfill
     \begin{subfigure}[b]{0.15\textwidth}
         \centering
    \includegraphics[width=\linewidth]{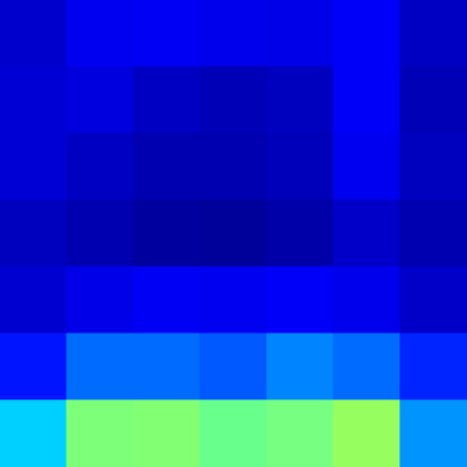}
    \caption{Wabot.a} 
     \end{subfigure}
    \caption{Example of cumulative HiResCAM heatmaps extracted from malware families with different CNNs.} 
    \label{fig:eg_heatmaps}
\end{figure*}

Note how the formula is the same as GradCAM, $\textbf{A}^f$ is again the feature map, but without the GAP multiplication step. The multiplication between the feature map and the gradients is done element-wise as indicated with $\odot$. Some examples of extracted heatmaps can be seen in Figure~\ref{fig:eg_heatmaps}.

\begin{figure*}
     \centering
     \begin{subfigure}[b]{0.15\textwidth}
         \centering
    \includegraphics[width=\linewidth]{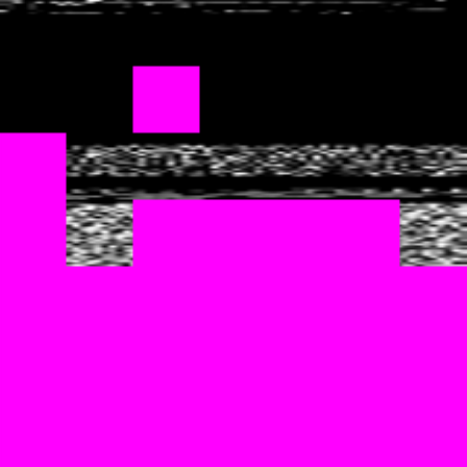}
    \caption{Agent.FYI} 
     \end{subfigure}
    \hfill
    \begin{subfigure}[b]{0.15\textwidth}         \centering
\centering
    \includegraphics[width=\linewidth]{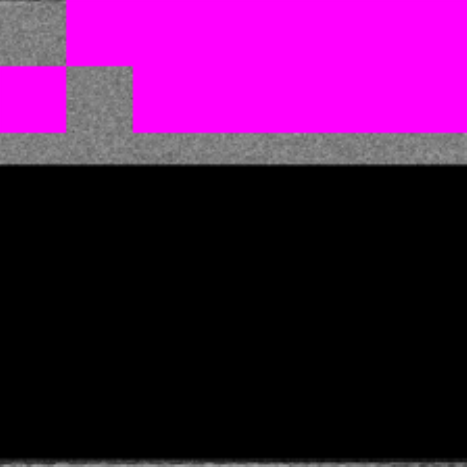}
    \caption{Khelios\_ver3} 
     \end{subfigure}
     \hfill
     \begin{subfigure}[b]{0.15\textwidth}
         \centering
    \includegraphics[width=\linewidth]{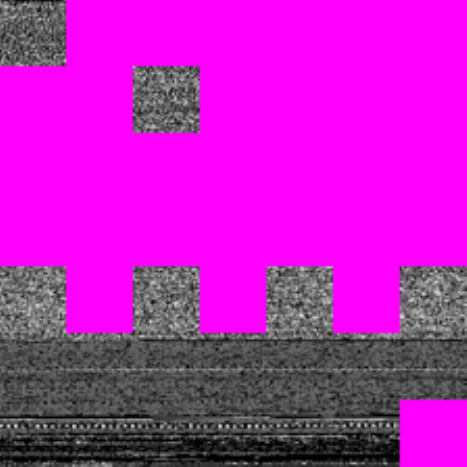}
    \caption{Rbot!gen} 
     \end{subfigure}
     \hfill
     \begin{subfigure}[b]{0.15\textwidth}
         \centering
    \includegraphics[width=\linewidth]{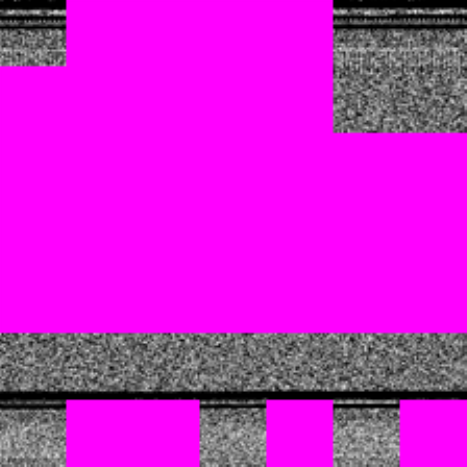}
    \caption{Skintrim.N} 
     \end{subfigure}
     \hfill
     \begin{subfigure}[b]{0.15\textwidth}
         \centering
    \includegraphics[width=\linewidth]{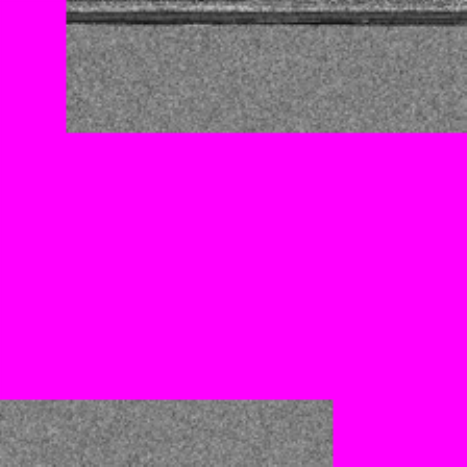}
    \caption{Wabot.a} 
     \end{subfigure}
    \caption{Masked samples for different malware families. The masks have been generated by merging \textit{EfficientNetB0} and \textit{DenseNet121} HiResCAM heatmaps.}
    \label{fig:maskedImages}
\end{figure*}
\subsubsection{Structural Similarity Index (SSIM)}
In our approach to comparing heatmaps and leveraging their information, we employ the Structural Similarity Index~(SSIM)~\cite{ssim}, a perception-based model designed for comparing images. 
SSIM values range from -1 to +1, reaching 1 only when two images are identical. This method was developed out of the necessity to identify similarities and differences within heatmaps. A mathematical model is incredibly important for accurately measuring differences. This is especially crucial when dealing with images, like malware images, which are difficult for humans to interpret. By using this method, it becomes possible to make precise and reliable assessments, overcoming the challenges posed by visual data that's hard to understand. We employed SSIM on cumulative heatmaps derived for each malware family within our datasets. These cumulative heatmaps, formed by adding and averaging the heatmaps of individual samples within a class, offer various valuable insights, mainly by enabling analysts to identify primary areas of focus when differentiating between one family and another during the classification process. The introduction of cumulative-SSIM allows researchers to assess both quantitatively and qualitatively the similarity between two cumulative heatmaps generated by distinct CNNs for the same family. 
We sought to extend the utility of heatmaps beyond their previously explained role as tools for elucidating CNN's performance. Our objective was twofold: firstly, to demonstrate that these heatmaps possess the potential to enhance classifiers' interpretability, and secondly, to explore their applicability in improving the overall performance of diverse architectures. We accomplished this by integrating the CNN-produced heatmaps into an ensemble method, thereby empowering classifiers to make more informed decisions. Through this approach, we aimed to showcase the ability of CNN guidance to enhance the decision-making processes of classifiers in effectiveness and interpretability. 

\subsubsection{Masks}

We introduce a novel approach similar to cutout regularization~\cite{cutout} involving the concept of a mask applied to malware image samples~\cite{wang2024malsort}. The purpose of this mask is to selectively cover specific regions of a malware image, rendering them inaccessible to the classifier during the analysis process. This strategic masking prevents the classifier from being misled by anomalous patterns occurring in positions where they typically do not appear and helps making sure the classifier is using only areas previously identified as important by other well performing techniques. This way, researchers can potentially use the same explanation developed for the starting model on the new model trained on masked images. The difference between our approach and the standard cutout regularization is that we do not use random patterns to cover areas of the image, but we specifically cover the areas that are identified by our explainability analysis as the areas with less useful information. In other words, the objective of the cutout is different from the concept of our masking. Cutout just removes random areas of the image to remove overfitting of a model in a dataset, while the masking applied in this study removes from the image specifically areas that might confuse the classifier.
Our approach to creating these masks combines two separate CNNs that independently generate heatmaps. These individual heatmaps merge into a unified heatmap using a logical \textit{OR} operation, forming the basis for defining the mask's coverage areas. The \textit{OR} operator is used because we want to keep all the information provided by both the CNNs. A heatmap is essentially a matrix with the same dimensions as the original sample. In this matrix, each pixel represents a value between 0 and 1 obtained through image normalization. A pixel with a value of 1 indicates consistent use by the CNN for predictions, while a value of 0 indicates non-use. After empirical observation, we found that the a 0.3 threshold consistently offered the desired precision level. We selected this threshold to exclude segments of the image that are rarely utilized. Pixels within the heatmap exhibiting values above this threshold indicate positions where the CNNs have identified relevant features essential for classification. In such instances, these pixels remain exposed, allowing the classifier to consider them during analysis. On the contrary, pixels with values below the 0.3 threshold are considered non-critical for classification. As a result, these pixels are concealed, ensuring that the classifier does not consider information from these regions. Through this masking technique, we mitigate the risk of classifier manipulation, thereby advancing the reliability of malware detection systems. Example of used masks can be seen in Figure~\ref{fig:maskedImages}.

\subsection{ViT Classifier}
In our research, we utilized the newly created masked datasets to enhance the classification efficacy of a Visual Transformer model~(ViT)~\cite{vit}. The inclusion of the ViT model in our study serves multiple purposes and addresses key methodological considerations. By incorporating ViT, we extend our analysis beyond CNNs, allowing us to evaluate our masking technique across fundamentally different approaches to image processing. This diversification helps avoid potential bias that could arise from testing CNN models on CNN-masked images, which might introduce a circularity in our methodology due to shared feature extraction mechanisms. One of our study's primary objectives is to highlight how different CNNs, despite employing similar techniques, may consider different areas as relevant. By introducing a non-CNN model like ViT, we create a valuable point of comparison, potentially revealing insights into the differences between CNN and transformer-based approaches in image interpretation. Using a model with a fundamentally different analysis method (self-attention instead of convolution) mitigates potential theoretical problems that might arise from testing conceptually similar models on each other's interpretations.
\par In computer vision, transformers can weigh the importance of different image regions by using self-attention, enabling them to capture complex spatial relationships and long-range dependencies. Unlike traditional attention mechanisms, where the attention weights are calculated based on the relationship between two separate sequences, self-attention allows a sequence to focus on different parts of itself. By considering all elements in the input sequence during processing, self-attention enables the model to access global contextual information. This is in contrast to traditional sequential models like RNNs, where information flow is limited by the fixed order of processing. Particularly against CNNs, researchers have found several advantages to using a ViT as compared to convolution models~\cite{vit4mal, vincent}. In particular, ViT is faster on edge devices due to the lack of convolution and pooling operations, it has less image-specific inductive bias than CNNs~\cite{vit}, it encodes information about the relative position of different spatial features and has local and global receptive fields, hence requiring smaller receptive fields for tracking long range dependencies.

The ViT implementation followed the framework provided in the Keras website\footnote{\href{https://keras.io/examples/vision/image\_classification\_with\_vision\_transformer/}{Keras ViT Implementation}, Accessed: 2023-5-15}, resulting in a minimalist architecture that incorporates a scaled-down version of the Visual Transformer initially introduced in~\cite{vit}. 
Even with the inherent simplicity of our proposed model, our research stands as proof of the potential these models hold when creatively applied, signaling, together with other recent research~\cite{vit4mal, vincent}, encouraging paths for future progress in the field.

\section{Experiments}
\label{sec:experiments}
In this section, we dive into our experiments aimed at answering the research questions introduced earlier. We start by explaining our experimental setup, including the tools and methods we used to replicate the selected models. We delve into the metrics employed for comparing our findings with those presented in the original papers. Finally, we introduce the technique involving visual transformers. By incorporating masks generated from heatmaps, we evaluate how this approach can enhance the performance of this architectural paradigm. We conducted the following experiments: tuning of the hyperparameters for the selected CNNs and replication of the model's training (\textbf{E1}), extraction of HiResCAM and GradCAM alongside the analysis of interpretability (\textbf{E2}), SSIM study on HiResCAM (\textbf{E3}) and the comparative study with ViT (\textbf{E4}).
\begin{table*}
\small
\centering
\begin{tabular}{|c|c|c|c|c|c|c|c|c|}
\hline
\multirow{2}{*}{\textbf{CNN}} & \multicolumn{2}{c|}{\textbf{Accuracy}} & \multicolumn{2}{c|}{\textbf{Precision}} & \multicolumn{2}{c|}{\textbf{Recall}} & \multicolumn{2}{c|}{\textbf{F1}} \\ \cline{2-9}

& \textbf{E} & \textbf{O}  & \textbf{E} & \textbf{O}  & \textbf{E} & \textbf{O}  & \textbf{E} & \textbf{O} \\
\hline
VGG16\cite{vasan2020imcec}       & 0.984 & 0.982 & \ & 0.957 & \ & 0.952 & \ & 0.954 \\
ResNet50\cite{vasan2020imcec}    & 0.982 &  0.986 & \ & 0.963 & \ & 0.962 & \ & 0.962 \\
IMCFN\cite{vasan2020imcfn}       & 0.978 & 0.982 & 0.982 & 0.954 & 0.981 & 0.952 & 0.982 & 0.953  \\
Gibert\cite{gibert2019using}      & 0.948 & 0.998 & \ & 0.965 & \ & 0.962 & 0.958 & 0.962 \\
EffNetB0\cite{pratama2022malw}  & 0.992 & 0.982 & 0.965 & 0.960 & 0.965 & 0.954 & 0.965 &  0.955 \\
DenseNet121\cite{tekerek2022anovel}     & 0.997 & 0.985 & 0.950 & 0.951 & 0.960 & 0.950 & 0.960 & 0.950 \\
\hline

\end{tabular}
\caption{models' performances. \textbf{E} represents the value expected found in the paper while \textbf{O} represents the value observed after replication.}

\label{tab:rep-malimg}
\end{table*}
\subsection{Replication (E1)}
\label{sec:51}
In this section we presented the environment and the metrics used to replicated the state-of-the-art models we selected. We will also detail the steps we used to make up for the missing information that we did not find in the papers, in particular the hyperparameters tuning step. Then, the results we obtained with our replicated models will be analyzed and explained.
\subsubsection{Environment}
All models were implemented in Keras\footnote{\href{https://github.com/fchollet/keras}{Keras}, Accessed: 2023-5-15} utilizing a Tensorflow backend\footnote{\href{https://www.tensorflow.org/}{TensorFlow}, Accessed: 2023-5-15}, in particular version 2.9.1. We had to use this Tensorflow version because of a known problem with the compatibility of the later version and the \textit{EfficientNetB0} model. The training and evaluation processes were conducted on the Google Colaboratory Pro+ plan, operating within a Python 3.7 environment. The runtime used in Colab was always with High-RAM enabled and T4 GPU Hardware accelerator. Data preprocessing, image creation, heatmap extractions, and all the scripting to facilitate the training of multiple neural networks with multiple datasets were carried out on an ASUS TUF Dash F15, running Windows 11. This system operates 16 GB of RAM and 12th Gen Intel Core i7. All our code and metrics calculated can be found on GitLab\footnote{\href{https://gitlab.com/mBRS_/reproducibility_explainability}{https://gitlab.com/mBRS\_/reproducibility\_explainability}}.

\subsubsection{Evaluation Metrics} 
In our study, we assessed model performance using accuracy, F1-score, recall and precision (see Equations~\ref{eq:acc}--\ref{eq:f1}). These evaluation metrics have been widely utilized in the research community to offer comprehensive assessments of various methods. In addition, we considered confusion matrices as another fundamental tool to investigate the architectures' performance. 

\begin{itemize}
   \item \textbf{True Positive~(TP)}: The model correctly identifies that a sample belongs to the correct class among multiple classes;
   \item \textbf{True Negative~(TN)}: The model correctly identifies that a sample does not belong to the wrong class among multiple classes;
   \item \textbf{False Positive~(FP)}: The model wrongly identifies that a sample belongs to the wrong class among multiple classes;
  \item \textbf{False Negative~(FN)}: The model wrongly identifies that a sample does not belong to the correct class among multiple classes.
\end{itemize}
The equations for each metrics are
\begin{equation}
\label{eq:acc}
\text{Accuracy} = \frac{\text{TP+TN}}{\text{TP+FN+TN+FP}}\text{,}
\end{equation}

\begin{equation}
\label{eq:precision}
\text{Precision} = \frac{\text{TP}}{\text{TP} + \text{FP}}\text{,}
\end{equation}

\begin{equation}
\label{eq:recall}
\text{Recall} = \frac{\text{TP}}{\text{TP} + \text{FN}}\text{,}
\end{equation}

\begin{equation}
\label{eq:f1}
\text{F1-Score} = 2 \times \frac{\text{Precision} \times \text{Recall}}{\text{Precision} + \text{Recall}}\text{.}
\end{equation}
\subsubsection{Hyperparameters Fine Tuning}
Once we had compiled the necessary information about the environment and settled on our hardware and software configurations, the next step involved delving into the specific architectures we intended to implement. We conducted a meticulous analysis of the papers and reached out to the authors in an attempt to obtain access to their source code. 
We then pinpointed the essential hyperparameters required for the implementations. The detailed values are presented in Table~\ref{tab:table-hyperparams2}. We adhered to the default settings provided by the Keras model for unspecified hyperparameters and we deviated from this approach only for hyperparameters specified in at least one paper. We developed a hyperparameter fine-tuning pipeline using Keras-tuner using the \texttt{RandomSearchCV()} technique\footnote{\href{https://github.com/keras-team/keras-tuner}{Keras Tuner}, Accessed: 2023-5-15}. 


The analysis of hyperparameter tuning results indicates a significant gap in the information provided by authors in the papers under review. The findings suggest that crucial details essential for replicating a model are often missing. Additionally, even when available, the provided information is frequently ambiguous, especially concerning aspects like the train-validation-test split. Consequently, researchers aiming to replicate a state-of-the-art malware image classification model face challenges as they need to adapt, infer, or fine-tune a considerable portion of the required hyperparameters on their own.

\subsubsection{Result Replication}

In Table~\ref{tab:rep-malimg} are presented the results of the replication study we did during this study. The models provided by~\cite{pratama2022malw} and~\cite{tekerek2022anovel} were tested on Big2015, while the models from~\cite{vasan2020imcec} and~\cite{vasan2020imcfn} were tested on Malimg. The architecture provided by~\cite{gibert2019using} was tested on both in the original paper but we considered only MalImg because for Big2015 the author used very different parameters when generating the image dataset. 
Our findings revealed disparities between the results obtained through implementation and those reported in the papers, with changes ranging from -0.029 to +0.004 when considering the F1 score. It is evident that the replicated models underperformed in comparison to the claims made in the original papers. This disparity underscores the challenge of replicating a model solely based on shared hyperparameters, even in cases where these parameters are widely available. Even though accurate replication is difficult due to the absence of both code and detailed architectural information, we managed to achieve results very similar to those reported in the papers, at least for some models. Other models didn't manage to obtain full compatibility of results, and even though we came close, in a field like malware classification where for the benchmark used the difference between old and stat-of-the-art is only a couple of \% points, small variations can make the difference between the optimal model and just a good model. This situation emphasizes the necessity for comprehensive code sharing as the sole viable solution for successful model reproduction.




A significant challenge during training preparation was the inconsistency in dataset splitting and testing methodologies. This issue was not unique to our study. To simulate a realistic approach, we opted for a 70/30 split for training and testing, reserving 10\% of the training set for validation.

After successfully replicating the results of the analyzed architectures, we optimized the models to ensure fair and consistent comparisons. Specifically, we adjusted the Gibert architecture~\cite{gibert2019using} by adding two convolutional layers, aligning the final layer's dimensions with the rest of the considered models. While this modification had a minor impact on the model's performance, it enabled meaningful comparisons when extracting heatmaps, as elaborated in the Section~\ref{sec:expl}.


\begin{table}
\small
\centering
\begin{tabular}{|c|c|c|c|c|}
\hline

\multirow{2}{*}{\textbf{CNN}} & \multicolumn{1}{c}{\textbf{Accuracy}} & \multicolumn{1}{c}{\textbf{Precision}} & \multicolumn{1}{c}{\textbf{Recall}} & \textbf{F1} \\ \cline{2-5}
& \multicolumn{4}{c|}{\textbf{Big2015}} \\\hline
VGG16 & 0.965 & 0.940 & 0.922 & 0.930 \\
ResNet50 & 0.979 & 0.959 & 0.944 & 0.951 \\
IMCFN & 0.973 & 0.954   &  0.946 & 0.950 \\
Gibert & 0.965 & 0.919 & 0.924 & 0.921 \\
EfficientNetB0 & 0.983 & 0.965 & 0.957 & 0.961 \\
DenseNet121 & 0.985 & 0.972 & 0.958 & \textbf{0.965} \\ \hline

& \multicolumn{4}{c|}{\textbf{MalImg}} \\
\hline
VGG16 & 0.982 & 0.957 & 0.952 & 0.954 \\
ResNet50 & 0.986 & 0.963 & 0.962 & 0.962 \\
IMCFN & 0.982 & 0.954  & 0.952 & 0.953 \\
Gibert & 0.975  & 0.922 & 0.925 & 0.923 \\
EfficientNetB0 & 0.990 & 0.974 & 0.973 & \textbf{0.973} \\
DenseNet121 & 0.984 & 0.963 & 0.956 & 0.956 \\ \hline

& \multicolumn{4}{c|}{\textbf{VX-Zoo}} \\
\hline
VGG16 & 0.945 & 0.911 & 0.909 & 0.910 \\
ResNet50 & 0.945 & 0.923 & 0.908 & 0.913 \\
IMCFN & 0.940 & 0.903  & 0.894 & 0.898 \\
Gibert & 0.950  & 0.920 & 0.915 & 0.916 \\
EfficientNetB0 & 0.969  & 0.953 & 0.959 & \textbf{0.955}  \\
DenseNet121 & 0.966 & 0.955  & 0.956 & \textbf{0.955} \\
\hline
\end{tabular}
\caption{Performance metrics of CNN models with split 70/30 on Big2015, MalImg and VX-Zoo}

\label{tab:7030-ogds}
\end{table}
The findings regarding replicability and reliability that can be found in Table~\ref{tab:7030-ogds} show how the metrics measured change, even significantly, when hyperparameters, in this case, training-validation-test split, are changed. Focusing on our two best models, \textit{DenseNet121} and \textit{EfficientNetB0}, a single change in this parameter resulted in a notable increase of +0.011 and +0.01 in the F1 score. These changes are substantial, considering that all classifiers under consideration exhibited precision values within the range of 0.1 to 0.15. We also tested all the models on our new dataset VX-Zoo, as seen in Table~\ref{tab:7030-ogds}. In this new dataset we can see how even the best model struggles to distinguish between classes \textit{Elkern.b}, \textit{Vtflooder.ekl} and \textit{Alman.b}, a situation analogous to MalImg, where two classes, in particular \textit{Swizzor.gen!E} and \textit{Swizzor.gen!I} cause the most trouble to researchers and their models~\cite{vasan2020imcec}\cite{vasan2020imcfn}\cite{gibert2019using}.

\subsection{Explainability}
\label{sec:expl}
\begin{figure*}
     \centering
     \begin{subfigure}[b]{0.15\textwidth}
         \centering
    \includegraphics[width=\linewidth]{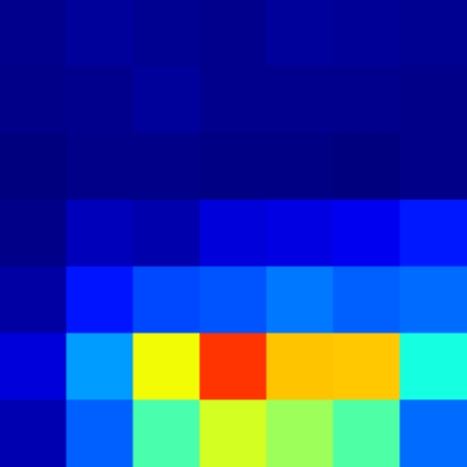}
    \caption{Vtflooder} 
     \end{subfigure}
    \hfill
    \begin{subfigure}[b]{0.15\textwidth}         \centering
\centering
    \includegraphics[width=\linewidth]{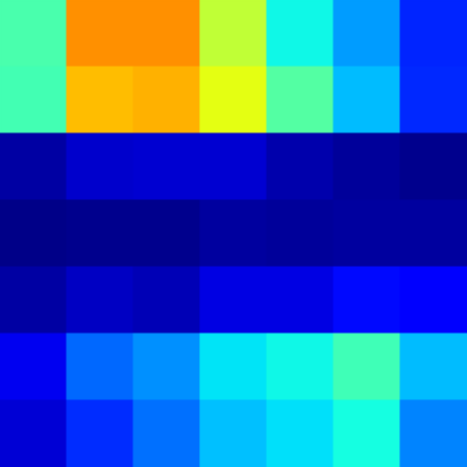}
    \caption{Elkern} 
     \end{subfigure}
     \hfill
     \begin{subfigure}[b]{0.15\textwidth}
         \centering
    \includegraphics[width=\linewidth]{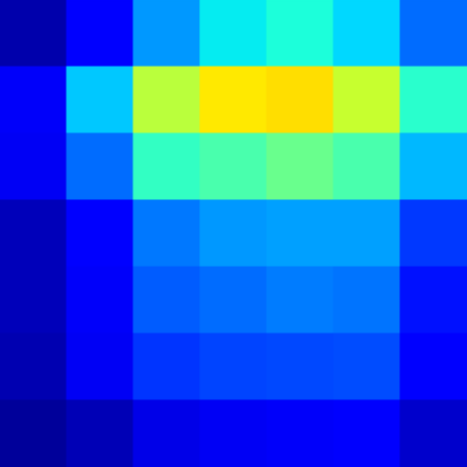}
    \caption{VB} 
     \end{subfigure}
     \hfill
     \begin{subfigure}[b]{0.15\textwidth}
         \centering
    \includegraphics[width=\linewidth]{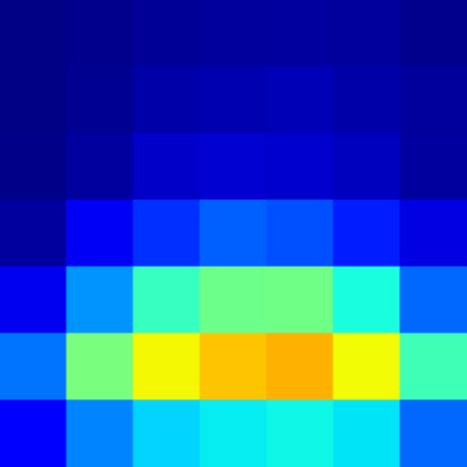}
    \caption{Expiro} 
     \end{subfigure}
     \hfill
     \begin{subfigure}[b]{0.15\textwidth}
         \centering
    \includegraphics[width=\linewidth]{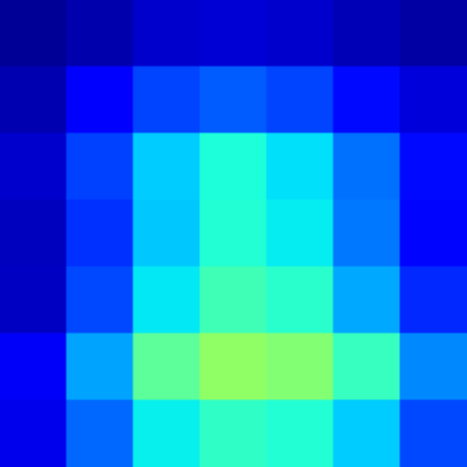}
    \caption{Parite} 
     \end{subfigure}
    \caption{Cumulative HiResCAM heatmaps extracted using \textit{ResNet50} on different malware classes of the VX-Zoo dataset.}
    \label{fig:otherMaps}
\end{figure*}
This section presents the explainability study applied to the models replicated in the previous parts. We describe the extraction of HiResCAM and GradCAM heatmaps for all the datasets and calculate the SSIM values for each family cumulative heatmaps. Then, we describe the deep learning model used as benchmark, namely the Visual Transformer, and its implementation. Finally, we introduce the original technique called masking that uses CAMs heatmaps to help the benchmark classifier obtain better results.

\subsubsection{Class Activation Maps Analysis (E2)}

We employed the earlier mentioned GradCAM and HiresCAM techniques on all the datasets available to extract cumulative heatmaps. Generating these heatmaps for each class was crucial for understanding whether diverse neural networks consistently utilized specific regions to classify particular samples. The heatmaps produced are $7 \times 7$ matrices that represent the feature map areas generated considering CNN's last convolutional layer. The resulting heatmaps can be found in the GitLab repository previously cited.

\begin{figure}
     \centering
     \begin{subfigure}[b]{0.45\textwidth}
         \centering
    \includegraphics[width=\linewidth]{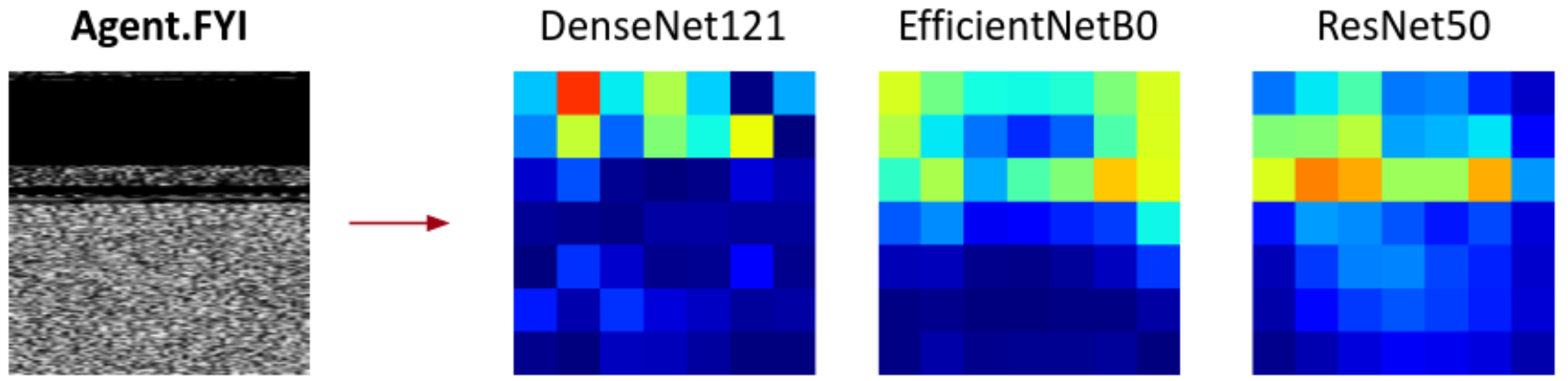}
    \caption{HiResCAM heatmaps produced for a single Agent.FYI sample.} 
     \end{subfigure}
     \par\bigskip 
     \begin{subfigure}[b]{0.45\textwidth}
         \centering
    \includegraphics[width=\linewidth]{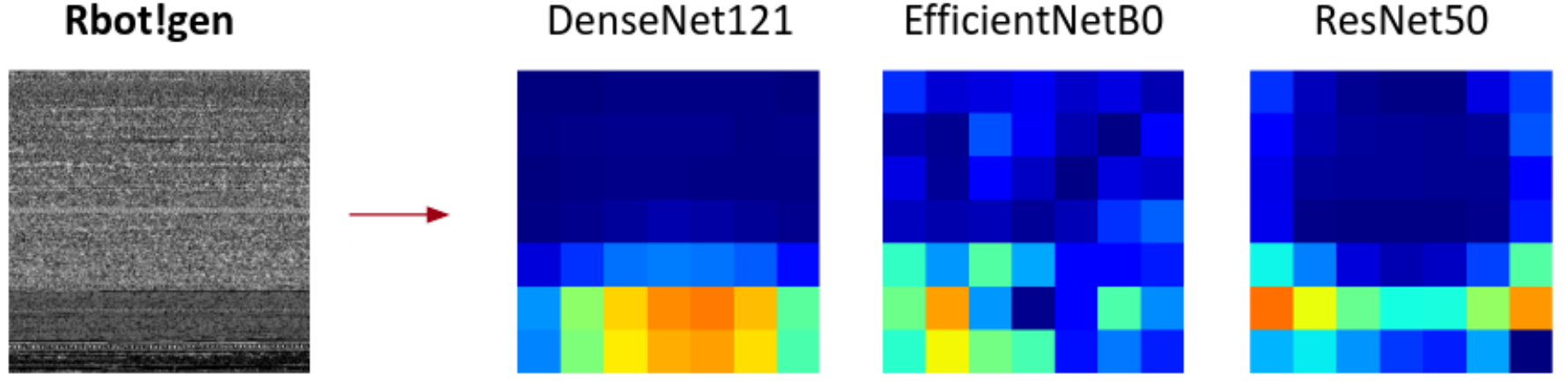}
    \caption{HiResCAM heatmaps produced for a single Rbot!gen sample.} 
     \end{subfigure}
     \par\bigskip 
     \begin{subfigure}[b]{0.45\textwidth}
         \centering
    \includegraphics[width=\linewidth]{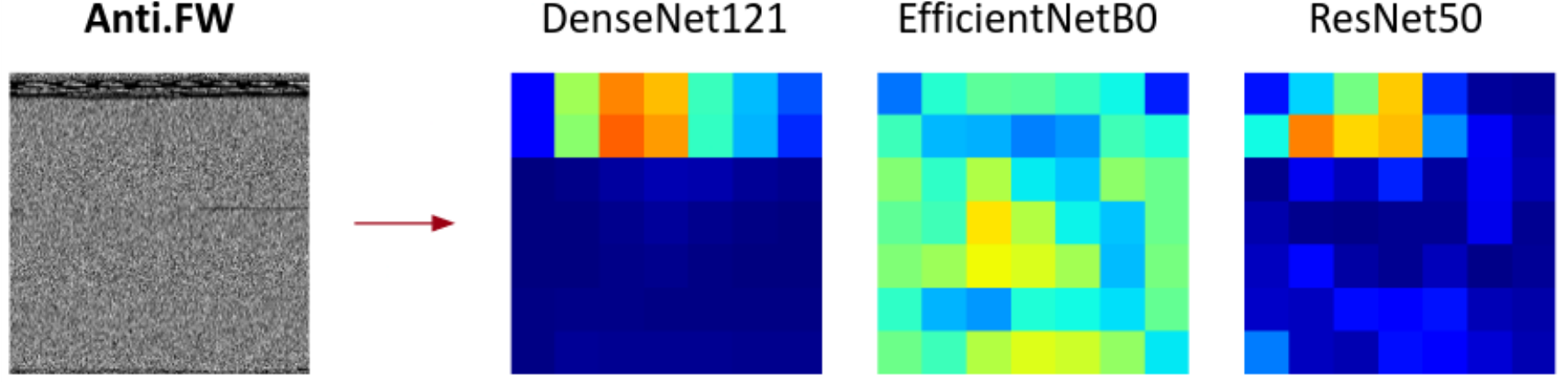}
    \caption{HiResCAM heatmaps produced for a single Anti.FW sample.} 
     \end{subfigure}
        \caption{Differences in areas analyzed by distinct CNNs. The heatmaps are extracted through HiResCAM. The use of different CNN can highlight the general consensus on a relevant region or the presence of different areas of interest considered by the models.}
        \label{agent_rbot}
\end{figure}
\begin{figure}
     \centering
     \begin{subfigure}[b]{0.49\linewidth}
     \centering
    \includegraphics[width=\linewidth]{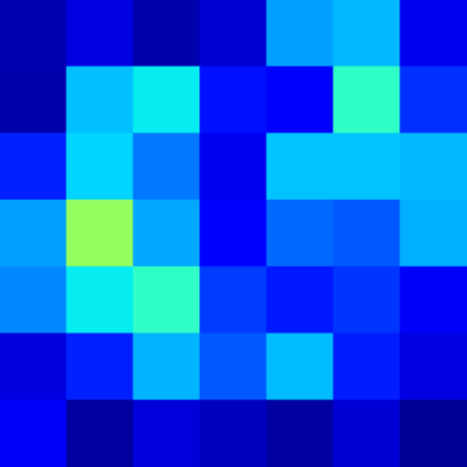}
     \caption{Swizzor.gen!E cumulative heatmap.} 
     \end{subfigure}
     \hfill
     \begin{subfigure}[b]{0.49\linewidth}
      \centering
    \includegraphics[width=\linewidth]{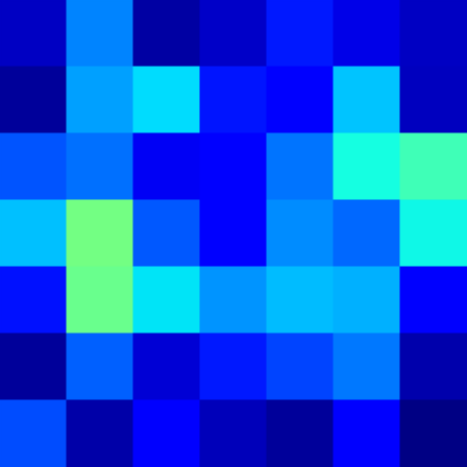}
    \caption{Swizzor.gen!I cumulative heatmap.} 
     \end{subfigure}
    \caption{Similarity between cumulative HiResCAM heatmaps extracted through \textit{DenseNet121} when applied to Swizzor trojan variants. Note how there are no clear hotspots, proving that the classifier struggles to identify salient characteristic in this type of malware.} 
        \label{swizzor}
\end{figure}

The analysis of heatmaps generated through HiResCAM, reveals discernible patterns, particularly within specific malware families. An intriguing observation arises when examining the operation of the same CNN across different classes within the same dataset.
Heatmaps can be used to compare the important features of different malware samples, helping analysts identify similarities or differences between variants of the same malware family. For example, if two samples have heatmaps that highlight similar regions, it may indicate shared code or functionality, similar obfuscation techniques or simply similar location of artifacts used for signatures. This can aid in clustering malware into families, tracking the evolution of malware over time, or identifying the reuse of code by threat actors. Different models can also prefer different areas, generating various possible artifacts locations or corroborating the same regions.
For instance, in the \textit{AntiFW.b} family, \textit{DenseNet121} and \textit{ResNet50} predominantly utilize the upper portion of the malware image to predict this specific class, indicating a consistent strategy. In contrast, \textit{EfficientNetB0} appears to employ a different approach for accurate classification within the same family. Another example is given by \textit{Rbot!gen.C} and \textit{Agent.FYI} classes from the MalImg dataset, where \textit{ResNet50}, \textit{DenseNet121} and \textit{EfficientNetB0} consistently directs attention to the same sections of the image as seen in Figure~\ref{agent_rbot}. From this results, and similar ones that emerge while studying other heatmaps, we can infer what are the patterns that the different CNNs are using to classify the samples. Even without returning to the binary form, we can identify in image areas, for example, encrypted sections, resources, or empty zones, that can direct the analyst to find a class signature or unique characteristic. This behavior highlights the classifier ability to concentrate its focus on a specific subset of an image, disregarding surrounding elements. This focused attention allows the network to accurately classify the sample by concentrating on the distinctive patterns unique to the specific family, thus showcasing the network's capacity to discern crucial features within localized sections of the input data.
Shifting focus to the previously mentioned \textit{Swizzor.gen!E} and \textit{Swizzor.gen!I} families, the cumulative heatmaps exhibit striking similarity as shown in Figure~\ref{swizzor}. The two classes are too similar to be distinguished by the CNNs and as a consequence the heatmaps do not identify clear salient point nor they provide a way to distinguish the two families. This situation is common when analyzing the \textit{Swizzor} macro family and is stressed by different authors~\cite{gibert2019using}\cite{nataraj2011malware}.

\par Another way HiResCAM heatmaps can be useful as a preliminary step for a human researcher is by identifying interesting areas of the malware. These areas can have two possible uses: defensively, a researcher can identify which areas contain possible signatures to be used to identify the class studied, and offensively, an attacker can use the heatmap to pinpoint the region where to introduce noise in order to disrupt the classifier. The next discussion is based on heatmaps found in Figure~\ref{fig:otherMaps}.
\par The Elkern family analysis revealed a noteworthy pattern in the HiResCAM heatmap. Two distinct macro areas were identified: one at the top, corresponding to the PE header region, and another near the file's end. Upon closer examination of the binary file, guided by the heatmap's indications, we found that these areas align with specific file sections. The top region corresponds to the aforementioned PE header, while the lower area matches the resources section, which includes exports and the entry point. By focusing directly on the resource section, we find some interesting strings that identify a Delphi binary (DVCLAL, PACKAGEINFO). Together with the exports, header values and entry point, these values can be used as a prototype for a signature.

The Vflooder family presents a more challenging scenario for interpretation. The CNN-generated heatmap highlights an area that, upon static analysis of the binary, falls between two sections. This unusual pattern necessitates a more thorough investigation to identify the family's specific characteristics. However, the heatmap's focus, particularly the area just preceding the imports, provides a valuable starting point for deeper analysis. This region could potentially offer insights into the malware's functionality and behavior, guiding further research efforts to unravel the intricacies of the Vflooder family, which represents a peculiar type of malware that is used to flood cloud sandboxes like VirusTotal.

\begin{figure}
     \centering
     \begin{subfigure}[b]{0.45\textwidth}
         \centering
    \includegraphics[width=\linewidth]{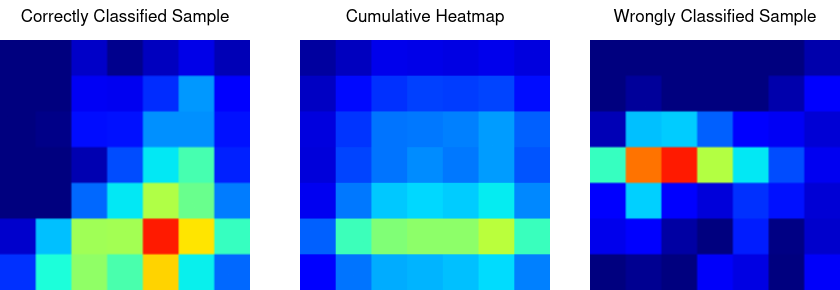}
    \caption{Correctly classified WBNA sample (left), cumulative heatmaps for the WBNA class (middle) and misclassified WBNA sample (right).} 
     \end{subfigure}
     \par\bigskip 
     \begin{subfigure}[b]{0.45\textwidth}
         \centering
    \includegraphics[width=\linewidth]{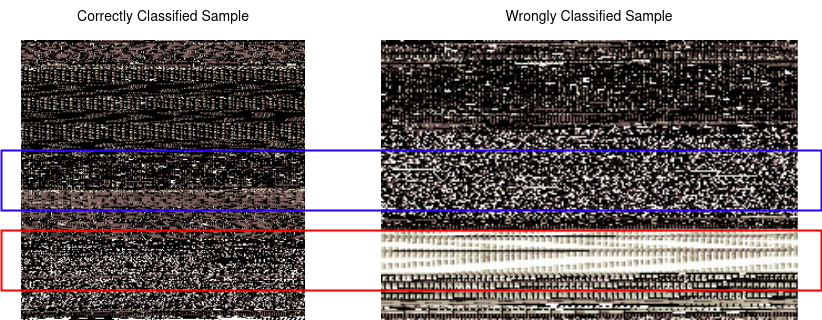}
    \caption{Samples for which CAM has been considered in the example.} 
     \end{subfigure}
        \caption{Comparison between a correctly classified and a misclassified sample both having as true class the \textit{WBNA} family. The cumulative heatmap shows the areas used by the CNN to make the classification. In subfigure (b) is highlighted in red the region usually activated for the \textit{WBNA} family, while in blue is highlighted the area activated for the wrongly classified sample. The red segment in the misclassified sample contains a peculiar pattern not usually found in other samples of the family, which is fooling the model.}
        \label{wbna_example}
\end{figure}

\begin{figure*}[h!]
\centering
    \includegraphics[scale=0.4]{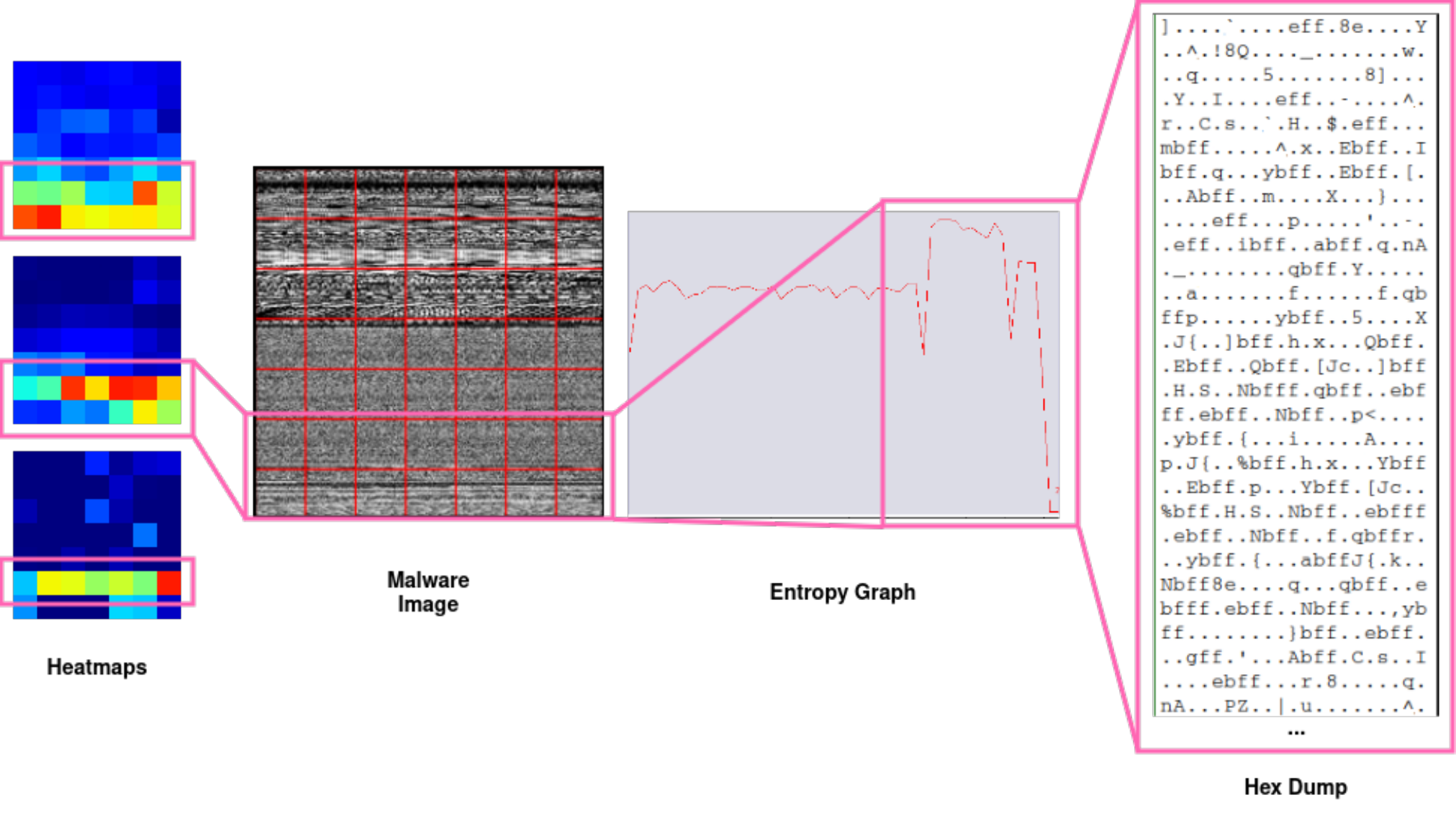}
        \caption{Process used to explain an Elkern sample by using a HiResCAM heatmap. The three heatmaps used an example are extracted using \textit{EfficientNetB0}, \textit{ResNet50} and \textit{DenseNet121} while the entropy graph shown is only for the \textit{.UPX} section of the sample.}
        \label{fig:extraction}
\end{figure*}

As another example, we used CNN heatmaps to identify the most important area for the class Expiro. The CNNs have identified a signature area—a zone that can be used to identify the class. If we statically analyze the code, we can see that this area is in the \textit{.UPX} section (packed) and, in particular, it occupies the ending. By looking at the entropy, we can see that this area has a higher entropy than the rest of the \textit{.UPX} section, which indicates that a section containing resources has been packed and that those resources can be used to identify the malware, even if still packed. This intuition given by the CNN made us skip all the cumbersome static analysis processes and directed us to a characteristic peculiar to this class. The heatmaps identify, with particular intensity, the area just before the presence of high entropy. By doing a hex dump and transforming the value in the ASCII character, we can see how a peculiar pattern appears; see in Figure~\ref{fig:extraction} the repetition of the \textit{ff} characters. This type of non-random repetition is of absolute interest for a malware analysis and highlights a meaningful and specific signature for the malware class.
Another use of saliency maps is in the analysis of misclassified samples. After identifying misclassified samples, we can generate feature importance maps and extract patterns across these samples. The class activation maps can be analyzed sample by sample, for example, by comparing the sample heatmap with the cumulative CAM heatmap or considering all the wrongly classified samples. These patterns will be analyzed to understand what regions or features consistently lead to wrong classifications. The specific outcome of these two types of analysis of misclassified samples depends heavily on the samples themselves, classes and general dataset structure, so it cannot be completely automated, but it will still facilitate the human malware analyst in his task. An example in our dataset is with the WBNA class, where a sample, Figure \ref{wbna_example}, is being misclassified by the \textit{ResNet50} CNN. Considering the cumulative heatmap of the WBNA class, we can see that it highlights very different areas than the average sample, suggesting where to look for the malware analyst. In the specific case provided in the example, the text section that is characteristic of WBNA has been significantly altered.
This type of study is also particularly important to validate whether the CNN is focusing on meaningful or suspicious regions of the binary image. For example, if the heatmap highlights areas corresponding to known malicious code sections (e.g., payloads or encryption routines), the analyst can confirm that the model is making logical decisions. Conversely, if the heatmap highlights irrelevant or benign sections, the analyst can investigate potential flaws in the model or training data, ensuring the CNN’s decisions are trustworthy and reliable.
The application of explainability tools as just shown serves a dual purpose. Primarily, the generated heatmaps provide valuable guidance to researchers, corroborating existing findings—such as those related to entropy—and directing attention towards the distinctive characteristics of specific malware strains. Moreover, and perhaps more significantly, these tools elucidate the rationale behind CNN's decision-making process. In the case study presented, the model's classification is attributed to the presence of particular patterns within a specific memory region. This transparency enables human researchers to critically assess the model's explanations, facilitating a qualitative evaluation that extends beyond mere performance metrics. Consequently, explainability introduces an additional dimension for assessing malware classifiers, encompassing not only their quantitative performance but also the soundness and interpretability of their decision-making processes. This multifaceted approach to evaluation enhances the overall reliability and trustworthiness of deep learning models in the domain of malware classification.

\subsubsection{SSIM Distance Evaluation (E3)}
We computed each family's cumulative Structural Similarity Index (cumulative-SSIM) value in every dataset. Initially, for each class within a dataset, a cumulative heatmap is generated by averaging all the heatmaps produced for each sample. Subsequently, the researcher computes the SSIM values between the heatmaps generated by different CNNs belonging to the same class. This process yields a specific value, denoted as single-class-SSIM, signifying how distinctively the analyzed CNN pairs interpret the same class. To obtain the cumulative-SSIM value between two CNNs, all the individual single-class-SSIM values are averaged. Each cumulative-SSIM value corresponds to a unique pair of CNNs and represents the disparity in their interpretations across all classes present in the dataset. Precise figures are provided for the VX-Zoo dataset in Table~\ref{tab:ssim-gradcam-cumul-ogds} and Table~\ref{tab:ssim-hirescam-cumul-ogds}, but a similar situation has been found in MalImg and Big2015. Elements in the tables are depicted in boldface to represent two CNN with high cumulative-SSIM and consequently high similarity, while elements in italics represent low cumulative-SSIM and consequently low similarity. We can see the difference in cumulative-SSIM produced by using two different methods on the same dataset and CNNs. 

\begin{table}[h!]
\scriptsize
\centering
\begin{tabular}{|c|c|c|c|c|c| }
\cline{1-2} 
EffNetB0 & \textit{1.5463}  \\ \cline{1-3}
Gibert & 2.169 & \textbf{\textit{1,617}}\\  \cline{1-4}
IMCFN & 1.852 & \textit{1.585} & 2,215  \\ \cline{1-5}
ResNet50 & \textbf{3.357} & \textit{1.405} & \textbf{2.394} & 1.828  \\\cline{1-6}
VGG16 & 1.975 & 1.81 & 2.206 & \textbf{3.446} & 1.762 \\\cline{1-6}
 & DenseNet121 & EffNetB0 & Gibert & IMCFN & ResNet50\\ \hline
\end{tabular}
\caption{GradCAM cumulative-SSIM values extracted from the VX-Zoo dataset.}
\label{tab:ssim-gradcam-cumul-ogds}
\end{table}
\begin{table}[h!]
\scriptsize
\centering
\begin{tabular}{|c|c|c|c|c|c|}
\cline{1-2} 
EffNetB0 & \textit{1.792} \\ \cline{1-3}
Gibert & 2.829 & \textbf{2.31} \\ \cline{1-4}
IMCFN & 2.152 & \textit{1.921} & \textbf{3.04}\\ \cline{1-5}
ResNet50 & \textbf{2.925} & \textit{1.344} & \textit{2.688} & 1.975  \\ \cline{1-6}
VGG16 & 2.121 & 2.219 & \textbf{3.262} & 2.962 & 1.891 \\ \cline{1-6}
&DenseNet121 & EffNetB0 & Gibert & IMCFN & ResNet50 \\ \hline
\end{tabular}
\caption{HiResCAM cumulative-SSIM values extracted from the VX-Zoo dataset.}
\label{tab:ssim-hirescam-cumul-ogds}
\end{table}
Greater similarity between the heatmaps generated by two CNNs, conveyed by the cumulative-SSIM, signifying a shared interpretation of malware families. Conversely, a lower value signifies divergent interpretations between the heatmaps, indicating dissimilar comprehension of the malware families.

Another way SSIM can be used is to qualitatively evaluate CNNs. We can calculate, as shown in Table~\ref{tab:ssim-best-cnn}, the average SSIM value obtained by confronting with each other all the class heatmaps extracted by a CNN. Applying this method, we obtained an SSIM value for each of the better performing CNNs tested on the 3 datasets. These SSIM values  represent how much the heatmaps differ and, consequently, how different the areas of the images used by the CNN to identify each class are. Intuitively, the model should be more confident in its decisions when the distance between its own heatmaps is higher. 
As we specified before, a lower SSIM value means that the heatmaps are different from each other. These numbers demonstrate that for 2 datasets over 3, the \textit{EfficientNetB0} CNN can better distinguish between malware families than the other 2 models. Indeed, the HiResCAM applied to \textit{EfficientNetB0} (and thus, the information in the convolution layers) is able to produce more characteristic heatmap than the other models. Thanks to these results, we can add a qualitative judgment to CNNs' performances other than taking the quantitative results at face value from Table~\ref{tab:7030-ogds}. For Big2015, \textit{DenseNet121} is the better-performing CNN, but \textit{ResNet50} is more robust. For MalImg, there is a coincidence between the better-performing model and the most solid. In the case of VX-Zoo, even though we have two similarly performing CNNs, a researcher should prefer \textit{EfficientNetB0} for the result obtained  in the SSIM study.

\begin{table}[h!]
\small
\centering
\begin{tabular}{|c|c|c|c|}
\hline
\textbf{Model}  & \textbf{Big2015}  & \text{MalImg}   & \textbf{VX-Zoo}   \\ \hline
EfficientNetB0  & 0.60050           & \textbf{0.48280}  & \textbf{0.52602}  \\ \hline
DenseNet121     & 0.61006           & 0.58666           & 0.59729           \\ \hline
ResNet50        & \textbf{0.54135}  & 0.49236           & 0.57170           \\ \hline

\end{tabular}
\caption{The average cumulative SSIM computed for the best-performing models on the datasets. A low SSIM value indicates that the analyzed images, in this context cumulative HiResCAM maps, exhibit significant dissimilarity. Consequently, a model yielding a low SSIM value is utilizing diverse regions to distinguish between various malware families, thereby demonstrating greater robustness.}
\label{tab:ssim-best-cnn}
\end{table}

\subsection{Attention Ensemble Architecture (E4)}
\label{sec:53}
We study another architecture, namely the ViT, that recently appeared in the literature related to malware analysis~\cite{vit4mal} to probe the potential of an innovative technique that, starting from CAM heatmaps might be a future helpful tool for the malware analyst. The result of the ViT testing together with the prediction using the masks presented in this section, can be found in Table~\ref{tab:7030-ViT}. In the table, specifically the ``Datasets" column, ED means that the dataset has been masked with the masks generated through \textit{EfficientNetB0} and \textit{DenseNet121}, and RE means that the dataset has been masked with the masks generated through \textit{EfficientNetB0} and \textit{ResNet50}. The results with an ``*" represent the values calculated using GradCAM masks instead of HiResCAM masks. The \underline{underlined} values represent the baseline that we wanted to improve while the \textbf{bold} values represent the best obtained result for each dataset. The results for this step are presented in previously cited Table~\ref{tab:7030-ViT}.
\begin{table}[h!tbp]
\small
\centering
\begin{tabular}{|c|c|c|c|c|}
\hline
\multirow{2}{*}{\textbf{Dataset}} & \multicolumn{4}{c|}{\textbf{Metrics}} \\
\cline{2-5}
& \textbf{Accuracy} & \textbf{Precision} & \textbf{Recall} & \textbf{F1} \\
\hline
Big2015	& \underline{0.907} & 0.884 & 0.802 & \underline{0.803} \\\hline
Big2015 (ED) 	& \textbf{0.954} & 0.942 & 0.877 & \textbf{0.892} \\
Big2015 (RE)	& 0.939 & 0.927 & 0.830 & 0.833 \\ 
Big2015* (ED) & 0.911 & 0.860 & 0.795 & 0.808 \\
Big2015* (RE)	 & 0.921 & 0.908 & 0.822 & 0.836 \\ \hline
MalImg 	& \underline{0.972} & 0.940 & 0.929 & \underline{0.929} \\\hline
MalImg (ED) 	& \textbf{0.982} & 0.953 & 0.955 & \textbf{0.953} \\
MalImg (RE)	& 0.974 & 0.942 & 0.938 & 0.938 \\ 
MalImg* (ED) & 0.966  & 0.931 & 0.920 & 0.923 \\
MalImg* (RE)	 & 0.966 & 0.936 & 0.917 & 0.920 \\ \hline
VX-Zoo 		& \underline{0.927} & 0.877 & 0.868 & \underline{0.871} \\\hline
VX-Zoo (ED) 	& 0.930 & 0.882 & 0.888 & 0.876 \\
VX-Zoo (RE) 	& \textbf{0.933} & 0.905 & 0.890 & \textbf{0.892} \\
VX-Zoo* (ED)  & 0.921 & 0.873 & 0.854 & 0.857 \\
VX-Zoo* (RE)	 & 0.925 & 0.870 & 0.877 & 0.871 \\ \hline
\end{tabular}
\caption{Performance metrics of ViT with and without masking.  The "*" represents the use of GradCAM.}
\label{tab:7030-ViT}
\end{table}

In the particular implementation, we decided to use combined heatmaps generated by couples of CNNs, in particular, \textit{EfficientNetB0}+\textit{DenseNet121}, and \textit{EfficientNetB0}+\textit{ResNet50}. The first pair was chosen because those two CNNs were the best performing on average across the three datasets. The second couple was selected because the HiResCAM cumulative-SSIM between them was the lowest in two of the three datasets, and in particular for VX-Zoo, as seen in Table~\ref{tab:ssim-hirescam-cumul-ogds}. 
We used a pair couple because a lower cumulative-SSIM means a greater difference between the heatmaps. Consequently, fusing two masks, the areas of the images the ViT uses to classify the malware will be larger. 
In short, we want to suggest what areas to focus on, while keeping a reasonable size of the image intact. The masking process is particularly delicate because we want that all the areas of the image that are not relevant being covered and all the useful parts being left as they are. 
In Figure~\ref{fig:maskedImages}, we illustrate how the masking algorithm operates on different samples. In the case of the \textit{Agent} sample, the algorithm preserves the top black pattern, while for \textit{Khelios\_Ver3}, it covers the top part while blacking out the central/bottom section. Examining other samples reveals distinctive patterns within the mask holes, indicating the technique's attempt to facilitate the classifier by concealing specific image textures while showing others.
Our analysis reveals that HiResCAM masks more effectively pinpoint the image regions deemed relevant by the CNNs, resulting in improved overall outcomes, as shown in Table~\ref{tab:7030-ViT}. In contrast, GradCAM produces broader, less focused heatmaps that don't tightly highlight the critical areas. This distinction is particularly significant in the context of malware images, which, unlike conventional photographs, are not easily interpretable by humans. The lack of precision in GradCAM's output can pose challenges for researchers attempting to identify specific malware signatures. These findings, combined with our previous results, conclusively demonstrate that HiResCAM is a superior CAM algorithm for explaining malware images compared to GradCAM.
The results in Table~\ref{tab:7030-ViT} demonstrate a noticeable enhancement achieved through our masking approach. Specifically, all combinations of applied masks outperform the baseline. The baseline is considered as the metrics calculated on the ViT results applied on the selected dataset without masking.

The findings consistently demonstrate that ViT, when trained with HiResCAM heatmaps, either matches or surpasses its accuracy performance with GradCAM masks. This outcome supports the initial hypothesis that heatmaps generated by HiResCAM contribute to more reliable results in ViT classification across the selected datasets. The analysis highlights the general efficacy of HiResCAM, proving our initial hypothesis that more faithful heatmaps produce generally better results.

\section{Conclusions}
\label{sec:conclusion}
This paper delves into two crucial aspects: replicability and explainability, crucial for the advancement of deep learning in malware image classification. Researchers need to prioritize consolidating existing state-of-the-art models while developing new architectures and presenting results. Ensuring sufficient information for replication by others is essential for scientific integrity. Challenges in reproducibility and replicability are multifaceted, requiring experimentation with diverse datasets and clarity in data division. Disclosure of hyperparameters enhances transparency and evaluation, and is mandatory to guarantee replicability by other research teams. 

Our study on explainability in black box models, particularly CNNs, emphasizes the importance of tool choice. HiResCAM heatmaps provide insights into CNN operations, highlighting differences between malware families and potentially identifying new malware signatures. By utilizing the SSIM metric, we demonstrated our ability to quantify the robustness of models, facilitating an informed selection of the most suitable CNN. Specifically, we highlighted that a higher-performing CNN may not always be the optimal choice when considering explainability. We also were able to use the heatmaps to identifies interesting areas of the binary from which malware researchers can start their study of the specific malware family using three classes extracted from our newly introduced dataset. We improved a baseline ViT malware classification model with an innovative technique called masking obtaining an increase of accuracy upwards to 5\% accuracy based on the dataset. As for future work, a more competitive implementation of Vision Transformer~(ViT) warrants exploration. Our masking technique shows promise in enhancing malware classifiers, emphasizing its role not only in getting better performance but also on explainability. Investigating whether masking datasets during training, employing our proposed method, leads to qualitative improvements and quantifiable enhancements in results, which is an area ripe for further study. Another important contribution would be testing on different types of datasets, like testing on only modern malware or on a larger number of classes. Even thought VX-Zoo in its current version is an improvement over other benchmark dataset, to obtain an analysis as close to the real world as possible, other datasets with different characteristics are needed.
\section*{Acknowledgment}
Authors’ contributions:
\textbf{Matteo Brosolo} -- conceptualization, data collection, dataset validation, software, experimentation, result analysis, writing.
\textbf{Vinod P.} -- conceptualization, dataset validation, experimentation, result analysis, writing, supervision.
\textbf{Mauro Conti} -- conceptualization, result analysis, writing, supervision.
\section*{Competing interests}
The authors declare that they have no conflict of
interest. The authors have no affiliation with or involvement in any organization or entity with a direct or indirect financial interest or non-financial interest in the subject matter discussed in the manuscript.
\section*{Funding sources}
\setlength{\intextsep}{0pt} 
\setlength{\columnsep}{0pt} 

\begin{figure}[!htbp]
\centering
    \includegraphics[scale=0.3]{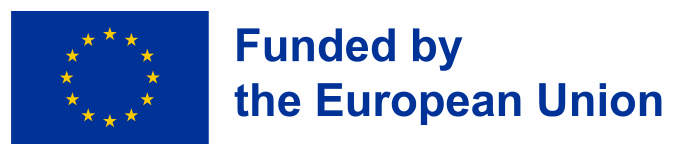}  
   \label{fig:eu}
\end{figure}
This work was partly supported by the HORIZON Europe Framework Programme through the project ``OPTIMA-Organization sPecific Threat Intelligence Mining and sharing” (101063107), funded by the European Commission . Views and opinions expressed are however those of the author(s) only and do not necessarily reflect those of the European Union. Neither the European Union nor the granting authority can be held responsible for them.

\bibliographystyle{elsarticle-num-names.bst} 
\bibliography{cas-refs.bib}





\end{document}